\newif\iflncs
\newif\ifnotes

\newif\ifblind\blindfalse

\notesfalse
\lncsfalse

\iflncs

\RequirePackage{amsmath}
\documentclass[runningheads]{llncs}
\titlerunning{}
\usepackage{orcidlink}
\usepackage{template_lncs}
\usepackage{global}

\else
\documentclass[12pt]{article}
\usepackage{times}
\usepackage{alpha_nat}
\usepackage{physics}
\usepackage{template}
\usepackage{global}
\fi

\usetikzlibrary{arrows.meta,positioning,calc}

\begin{document}

\title{Limitations on Autoregressive Token Generation via Sequential Sampling}
\author{%
Gal Arnon \thanks{\texttt{galarnon42@gmail.com}. The majority of this work was done while the author was a postdoc at Bocconi University} \\ NTT Research
\and Andrej Bogdanov \thanks{\texttt{abogdano@uottawa.ca}} \\ University of Ottawa
\and Paul Lou \thanks{\texttt{paul96lou@gmail.com}} \\ Bocconi University
\and Giulio Malavolta \thanks{\texttt{giulio.malavolta@hotmail.it}} \\ Bocconi University
\and Alon Rosen \thanks{\texttt{alon.rosen@unibocconi.it}} \\ Bocconi University
\and Lorenzo Taschin \thanks{\texttt{taschin1998@gmail.com}} \\ University of Ottawa
}

\iflncs

\else

\fi
\maketitle

\begin{abstract}
    We study autoregressive token generation through the notion of  sequential samplers: A circuit that samples the next symbol only given the previous ones and possibly a small amount of state. 
    We show that (1) a sequential sampler with $k$ bits of state can be simulated by a stateless one at a roughly $O(2^{k})$ size overhead; (2) efficient insertional sampling (Stern, Chan, Kiros, Uszkoreit, ICML 2019) is more powerful than efficient sequential sampling in any fixed order; (3) sequential samplers with small state cannot establish certain forms of covert cryptographic channels.  Result (1), together with the pseudoentropy vs. sequential sampleability duality of Vadhan and Zheng (STOC 2012),
    characterizes efficient sequential sampleability with small state.  Result (3) provides a potential safety mechanism against subliminal coordination among malicious LLM agents.
\end{abstract}

\newpage
\tableofcontents
\newpage

\iflncs

\else

\section{Introduction}

Autoregressive language models generate an entire response through a
sequence of local decisions: sample the next token given the prompt and
the tokens already produced. An ordinary randomized sampler, in contrast, is unconstrained by the order in which its output is produced. 
A long pseudorandom string, for example, can be derived from a short random seed, even though the visible prefix does
not reveal its continuation.

In this work we fully characterize the class of distributions that can be sampled \emph{efficiently} one token at a time, using only the preceding output.

The probability chain rule factors \emph{every} distribution into next-token conditionals: $p(x_1x_2x_3) = p(x_1)p(x_2|x_1)p(x_3|x_1x_2)$. However, this formula says nothing about the complexity of sampling those conditionals. An efficient algorithm for generating a whole string need not provide an efficient way to continue from a prefix drawn from the same distribution.  \citet{ImpagliazzoLevin1990} furnish a sufficient hypothesis for \emph{universal} efficient sequential sampleability:  The average-case invertibility of all efficient functions.  Their hypothesis is unreasonable as it would forbid cryptography.  (We may be in for surprises!)

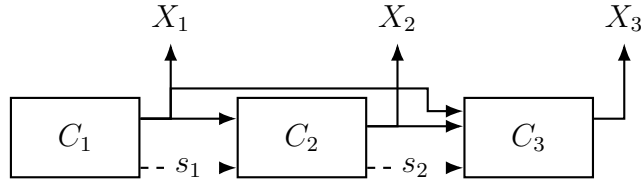
\begin{figure}[ht]
\centering
\begin{tikzpicture}[
    >=Latex,
    block/.style={
        draw,
        thick,
        minimum width=1.7cm,
        minimum height=1.05cm,
        align=center
    },
    signal/.style={
        ->,
        thick
    }
]

\node[block] (C1) at (0,0)   {$C_1$};
\node[block] (C2) at (3,0) {$C_2$};
\node[block] (C3) at (6,0) {$C_3$};

\coordinate (C12) at ([yshift=2.5mm]C1.east);
\coordinate (C21) at ([yshift=2.5mm]C2.west);

\coordinate (C23) at ([yshift=1.5mm]C2.east);
\coordinate (C32) at ([yshift=1.5mm]C3.west);

\coordinate (C12s) at ([yshift=-4mm]C1.east);
\coordinate (C21s) at ([yshift=-4mm]C2.west);

\coordinate (C23s) at ([yshift=-4mm]C2.east);
\coordinate (C32s) at ([yshift=-4mm]C3.west);

\draw[signal] (C12) -- (C21);
\draw[signal] (C23) -- (C32);

\draw[signal, dashed] (C12s) -- node[fill=white] {$s_1$} (C21s);

\draw[signal, dashed] (C23s) -- node[fill=white] {$s_2$} (C32s);

\coordinate (C3out) at ([yshift=2.5mm]C3.east);

\draw[signal] (C12) -- ++(0.4, 0) -- ++(0, 1) node[above] {$X_1$};

\draw[signal] (C12) -- ++(0.4, 0) -- ++(0, .4) -- ++ (3.4, 0) -- ++ (0, -.3) -- ([yshift=3.5mm]C3.west); 

\draw[signal] (C23) -- ++(0.4, 0) -- ++(0, 1.1) node[above] {$X_2$};

\draw[signal] (C3out) -- ++(0.4, 0) -- ++(0, 1) node[above] {$X_3$};

\end{tikzpicture}
\caption{A three-token sequential sampler.}
\label{fig:sampler}
\end{figure}

We study an abstraction of autoregressive generation called a
\emph{sequential sampler} (see Fig.~\ref{fig:sampler}).  This is a collection of probabilistic circuits $C(1^n) = (C_1, C_2, \dots, C_n)$ generating token sequence $X_1, X_2, \dots, X_n$.  Circuit $C_i$ takes as inputs the prefix $X_{<i} = (X_1, \ldots, X_{i-1})$, and possibly a bounded state $s_{i-1}$ produced by $C_{i-1}$.  It outputs the next token-state pair $(X_i, s_i)$.   When $s_i$ are empty the sampler is  \emph{stateless}.

Every step can reread the entire prefix sampled so far.
Private information generated during one step, including randomness, can only be explicitly passed along through the state.
Prompts, as well as other fixed model parameters, can be
hardwired into the circuits, so we impose no restriction on the training procedure, attention mechanism, or depth of the next-token computation beyond efficiency, i.e., that the circuit size be polynomial in the total output length.  The model captures generation from fixed autoregressive parameters when all information retained between steps is determined by the prompt and generated prefix.

Earlier work by \citet{lin2021limitations} identifies sampleability limitations for autoregressive models satisfying the stronger assumption that next-token probabilities are efficiently computable. \citet{fortnow2026} shows that the distributions that can be sampled in this manner are precisely the ones whose \emph{cumulative} distribution function is efficiently computable~\cite{levin1986}.  Our model allows for a richer class of distributions.

More relevant to our model is the limitation observed by \citet{wu2023limitations}: A stateless sequential sampler cannot generate a number followed by its prime factorization, unless factoring is easy.  \citet{vadhan-zheng} and \citet{C:ACHV19} fully characterize this type of obstruction for stateless, fixed-order samplers.

Our approach is based on the duality between conditional sampleability and \emph{pseudoentropy} discovered by 
\citet{vadhan-zheng}.  \citet{C:ACHV19} interpret this connection in the context of stateless sequential sampling.  We derive analogous characterizations for stateful and \emph{insertional} samplers.

\subsection{Our Results}
All reasonable sampling notions tolerate some error of the sampler $C(1^n)$ relative to the target distribution $Y$. A natural measure, which we adopt in this work, is their KL-divergence $\kldiv{Y}{C(1^n)}$. Note that small KL-divergence implies small total variation distance.

\paragraph{Characterizing sequential samplers.} For left-to-right generation, the relevant quantity is
\emph{next-block pseudoentropy}. It measures how much entropy the next
token can appear to have, given the true preceding prefix, summed over
the positions. The \emph{next-block pseudoentropy gap} is the difference
between this quantity and the true entropy of the complete string. This equivalence follows from the work of \citet{vadhan-zheng} and was informally discussed by \citet{C:ACHV19}. We give a formal, quantitative statement and proof:

\begin{theorem}[\citet{vadhan-zheng, C:ACHV19}, see \Cref{thm:sequential-characterization}]
\label{thm:intro-sequential}
A distribution has a small next-block pseudoentropy gap if and only if
it admits an efficient stateless sequential sampler with small KL error.
\end{theorem}

The next-token circuits can be chosen with full
knowledge of the target, independently of any learning algorithm. A large pseudoentropy gap therefore identifies a limitation that persists
even when training and model selection are unrestricted.

We show that the above extends beyond stateless samplers.  The reason is that private state can be eliminated, at a cost polynomial in the number of possible states:
\begin{theorem}[\Cref{thm:state-elimination-KL}, informal]
\label{thm:intro-state-elimination}
A sequential sampler with $k$ bits of state can be approximated in KL-divergence by a stateless sequential sampler. The circuit-size overhead is $2^{k}$ and polynomial in the other parameters, including token alphabet size.  
\end{theorem}

A factor $2^k$ blowup in overhead is inevitable---in the worst case---for simulation that estimates the state distribution from independent samples.  As intuition, a pseudorandom generator with a $k$-bit seed admits an efficient stateful implementation that stores the seed.  A stateless implementation, however, seems to effectively require brute-force search over all possible seeds. See \Cref{subsec:state-distribution-lower-bound}.

\paragraph{Insertional Samplers.}  An \emph{insertional sampler} can insert a new token anywhere in the output, not just at the terminal position.  The Insertion Transformer~\cite{DBLP:conf/icml/SternCKU19} is endowed with this capability.  The sampler still retains bounded private state, but can generate information in an order suited to the task. 

We introduce the notion of \emph{insertional
pseudoentropy gap} to measure how much the apparent conditional entropy can exceed the actual conditional entropy, summed over the insertions, for every insertional decomposition of the target.  We show a duality between insertional pseudoentropy and sampleability:

\begin{theorem}[\Cref{thm:insertional-characterization}, informal]
\label{thm:intro-insertional}
A distribution has a small insertional pseudoentropy gap if and only if
it admits an efficient stateless insertional sampler with small KL error.
\end{theorem}

Insertional samplers are more powerful than sequential ones: Going back to the factoring example, one can simply output the factors first, then insert their product in the first position of the sequence. In fact, this is not just a property of factoring: For any one-way function $f$,\footnote{Recall that a one-way function $f$ is a function where $y=f(x)$ is easy to compute for every $x$, but it is hard to find a preimage of $y$ when $x$ is sampled uniformly  at random.} the pair $(f(X), X)$ is hard to sample in a left-to-right manner but easy to sample insertionally (\Cref{thm:insertional-vs-sequential}).

In contrast to the pair $(f(X), X)$ which is easy to sample from right-to-left, a pseudorandom generator\footnote{A pseudorandom generator $G \colon \zo^{n} \to \zo^{m}$ is a length-extending function where $G(s)$ for a uniformly random $s\gets \zo^{n}$ is computationally indistinguishable from a uniformly sampled string over $\zo^{m}$.} is not sequentially samplable for \emph{any fixed} ordering.
This is because it has next-bit pseudoentropy with respect to every fixed ordering of the output bits  (see \Cref{cor:no-stateless-prg}). Thus, one might expect that the pseudorandom generator output distribution is not insertionally sampleable. Yet, surprisingly, we demonstrate the existence of pseudorandom generators whose output can be sampled insertionally:
\begin{theorem}[\Cref{prop:insertionally-samplable-prg}, informal]
    Assuming one-way functions exist, there is a PRG whose distribution can be sampled by a stateless insertional sampler.
\end{theorem}
Our construction crucially exploits the stochasticity of the insertion positions. We further demonstrate a separation between insertional and arbitrary samplers, indicating a limitation of their computational capabilities:

\begin{theorem}[\Cref{thm:sm-insertional-separation}, informal]
Assuming the existence of very hard-to-invert one-way functions, there is an efficiently sampleable distribution that cannot be sampled efficiently in a stateless insertional manner.
\end{theorem}

The token alphabet in our hard distribution has bit-length linear in the sequence length $n$.  We do not know if this dependence is necessary.

\paragraph{Consequences (\Cref{sec:consequences}).} Our results reveal fundamental tradeoffs between the expressive power of generative models and their cryptographic capabilities.  More restrictive architectures provide automatic safety guarantees.  More expressive architectures, in contrast, could be useful for training cryptographic functionalities.

\smallskip\noindent\emph{Cryptographic capabilities and safeguards.}
    Autoregressive models face limits even on useful tasks that ordinary programs perform routinely. 
    Our results imply that they cannot faithfully sample the output of a pseudorandom generator, a necessity for encryption.  Supporting these capabilities requires sufficient private state or access to external resources. 
    
    This limitation can be beneficial for safety. 
    Untrusted agents might try to establish a covert channel via steganographic techniques.  Bounding the state of transformer-type agents constrains their ability to communicate in secret.

\smallskip\noindent\emph{A case for insertion-based models.}
    At the level
    of generation rules, insertion is strictly more expressive than
    left-to-right generation in our stateless setting. The freedom to choose
    where to write can therefore expand a model's capabilities without
    adding private memory. For example, as we have shown, useful cryptography can be computed statelessly by insertion-based models but not by models with any apriori fixed order. This provides a theoretical foundation for
    pursuing insertion-based models as an alternative to strictly
    autoregressive generation.

\subsection{Further Related Work}

Computational entropy provides several ways to express the difficulty
of producing correlated outputs. Inaccessible entropy, introduced by
\citet{STOC:HRVW09}, measures the entropy an online generator
can produce when its previous random coins are included in the
conditioning information. \citet{C:ACHV19} unify this
notion with next-block pseudoentropy through hardness in relative
entropy. We use the prefix-only sampling connection from this theory
as a foundation; the insertional analysis must additionally handle
intermediate strings and the information in their generation order.

Our next-token circuits sample their outputs, and their output
probabilities need not be efficiently evaluable. This distinguishes the
model from the probability-evaluation setting of
\citet{lin2021limitations}. Wu's factoring-based
example~\cite{wu2023limitations} already exhibits a limitation for
randomized prefix-only generation. The entropy characterization provides
a general criterion; our insertional analysis extends this viewpoint to
alternative generation orders.

Recent work also studies the computational role of generation order.
\citet{kim2025ordering} analyze hard masking conditionals under
computational assumptions and the benefits of choosing favorable orders.
\citet{yang2026powerful} compare autoregressive, masked, and
editing-based generation; their autoregressive simulation of arbitrary
orders emits position--token pairs and permits postprocessing. Our
samplers must directly produce the prescribed output string, so that
simulation does not imply equivalence in our model.
\citet{jiang2026parallel} study parallel sampling efficiency
and simulation of randomized circuits by diffusion models with auxiliary
tokens. Their resource bounds complement our focus on the distributions
that can be generated under restrictions on private state and output
order.

Mazor, Morgan, and Pass~\cite{MMP26} show that efficient autoregressive
models with efficiently evaluable next-token probabilities are equivalent
to efficient monotone, prefix-free compression, up to an additive two
bits between excess cross-entropy and compression redundancy. Combining
this with the Vadhan--Zheng characterization relates next-bit
pseudoentropy to monotone incompressibility. Their cryptographic
separation uses a distribution of the form $(G(U),U)$ for a
pseudorandom generator $G$, closely related to our fixed-order sampling
obstruction based on $(f(U),U)$. Our main results quantify the cost of
eliminating private state and characterize the power and limitations of
insertional sampling, allowing general randomized transitions without
requiring efficient probability evaluation. Thus, the two works share a
fixed-order pseudoentropy foundation while developing complementary
compression and sampling perspectives. Interestingly, the notion of insertional generation does not seem to be captured by monotone compression.

\section{Preliminaries}

{\sloppy 

For $n \in \bN$, define the set $[n] = \{1, \ldots, n\}$.
For any set $X$, we define $X^0 \coloneqq \{ \emptyString \}$ where $\emptyString$ denotes the empty string.
We say that a string $x$ is a subsequence of $y$ if $x$ can be produced from $y$ by deletions.  

For any two random variables or two distributions $X, Y$, we will use $\statDist(X, Y) \coloneqq \frac{1}{2} \cdot  \sum_{a \in \supp(X) \cup \supp(Y)} \abs{\Pr[X = a] - \Pr[Y =a] } \in [0, 1]$ to denote the total variation distance between $X$ and $Y$.
Our default logarithm function $\log$ is in base two.
Every randomized circuit uses fresh independent fair coins.
If $X, Y$ are random variables where $Y$ has support $\cY$, then for any $y \in \cY$ we use the notation $X \mid_{Y = y}$ to denote the conditional distribution of $X$ conditioned on the event that $Y = y$.

A function $f$ in parameter $\secParam \in \bN$ is noticeable if there exists a polynomial $p(\secParam)$ such that for all sufficiently large $\secParam$, $f(\secParam) > 1/p(\secParam)$.  
For a parameter $\secParam \in \bN$, for any two functions $\alpha = \alpha(\secParam)$ and $\beta = \beta(\secParam)$, we say that $\beta$ is noticeably larger $\alpha$ if there exists a noticeable function $\gamma = \gamma(\secParam)$ such that $\beta > \alpha + \gamma$.
A function $f$ in parameter $\secParam \in \bN$ is negligible if for all polynomials $p(\secParam)$, for sufficiently large enough $\secParam$, we have $f(\secParam) < 1/p(\secParam)$.
}
\subsection{Basic Information Theory}

We now recall relevant definitions and related facts from information theory.

\begin{definition}[Shannon entropy]
    For a discrete random variable $X$ with finite support, we define the Shannon entropy of $X$ to be $H(X) \coloneqq - \E_{x \sim X} [ \log_2 \Pr(X = x) ]$.
\end{definition}

\begin{definition}[Binary entropy]
For $p \in [0, 1]$, define the  binary entropy function $h(p) = - p \log_2 p - (1-p) \log_2 (1-p)$.
\end{definition}

\begin{definition}[KL divergence]
Let $P,Q$ be distributions over a finite set $\Omega$. If $P$ is absolutely
continuous with respect to $Q$, meaning that $P(a)>0$ implies $Q(a)>0$, then the
KL divergence from $P$ to $Q$ is
\[
D_{\mathrm{KL}}(P\|Q)
\coloneqq
\sum_{a\in\Omega} P(a)\log \frac{P(a)}{Q(a)}.
\]
If $P$ is not absolutely continuous with respect to $Q$, we define
$D_{\mathrm{KL}}(P\|Q)=+\infty$.
\end{definition}

\begin{fact}[Nonnegativity of KL divergence; see Theorem 2.6.3 in~\citet{cover-thomas}]
\label{fact:kl-nonnegative}
For any two distributions $P,Q$ over a finite set $\Omega$,
$D_{\mathrm{KL}}(P\|Q)\geq 0$.
\end{fact}

\begin{fact}[Pinsker's inequality; Lemma 11.6.1 in~\citet{cover-thomas}]
\label{fact:pinsker}
For finite distributions $P,Q$,
\[
\statDist(P,Q)\leq\sqrt{(\ln2)\kldiv{P}{Q}/2}.
\]
\end{fact}

\begin{definition}[Markov chain; see Chapter 4 in~\citet{cover-thomas}]
\label{def:markov-chain}
A sequence of random variables $Z_1,\ldots,Z_\outLen$ over finite sets is a Markov
chain, denoted $Z_1-Z_2-\cdots-Z_\outLen$
if for every $i\in\{2,\ldots,\outLen-1\}$,
\[
\Pr[Z_{i+1}=z_{i+1}\mid Z_1=z_1,\ldots,Z_i=z_i]
=
\Pr[Z_{i+1}=z_{i+1}\mid Z_i=z_i]
\]
whenever the conditioning event has positive probability.

Equivalently, for every $i\in\{2,\ldots,\outLen-1\}$,
$Z_{i+1}$ is conditionally independent of $(Z_1,\ldots,Z_{i-1})$ given $Z_i$.
\end{definition}
The KL chain rule compares two joint distributions by first comparing one
variable, then comparing the other conditioned on the first.

\begin{fact}[KL chain rule; Theorem~2.5.3 in~\citet{cover-thomas}]
\label{fact:kl-chain-rule}
Let $P_{XY}$ and $Q_{XY}$ be distributions over the same finite set
$\mathcal X\times\mathcal Y$. Then
\[
\kldiv{P_{XY}}{Q_{XY}}
=\kldiv{P_X}{Q_X}
+\E_{x\sim P_X}\kldiv{P_{Y\mid X=x}}{Q_{Y\mid X=x}}.
\]
Conditional distributions on events of probability zero may be chosen
arbitrarily, and the expectation sums only over $x$ with $P_X(x)>0$.
The identity allows infinite divergence.
\end{fact}

Iterating the chain rule gives a sum over steps, each conditioned on the
full preceding history. For Markov chains, the conditioning reduces to
the preceding variable.

\begin{fact}[KL chain rule for Markov chains]
\label{fact:kl-chain-rule-markov}
Let $n\geq1$ and let
$X_0,\ldots,X_n$ and $Y_0,\ldots,Y_n$ be Markov chains, where for each
$i$ the variables $X_i,Y_i$ take values in the same finite set, which may
depend on $i$. If they have the same initial distribution, $X_0\equiv Y_0$,
then
\begin{equation}
\label{eq:kl-chain-rule-markov}
\kldiv{(X_0,\ldots,X_n)}{(Y_0,\ldots,Y_n)}
=\sum_{i=1}^n\E_{x\sim X_{i-1}}
\kldiv{X_i\mid_{X_{i-1}=x}}{Y_i\mid_{Y_{i-1}=x}}.
\end{equation}
Both transitions are conditioned on the same value $x$, sampled from
$X_{i-1}$, with the same conventions for zero-probability events as above.
\end{fact}
\begin{proof}
Applying \Cref{fact:kl-chain-rule} to the first coordinate gives
\begin{align*}
&\kldiv{(X_0,\ldots,X_n)}{(Y_0,\ldots,Y_n)}\\
&\quad=\kldiv{X_0}{Y_0}
+\E_{x_0\sim X_0}
\kldiv{(X_1,\ldots,X_n)\mid_{X_0=x_0}}
       {(Y_1,\ldots,Y_n)\mid_{Y_0=x_0}}.
\end{align*}
Here $\kldiv{X_0}{Y_0}=0$ because $X_0\equiv Y_0$.
For $n=1$, this already gives the claim. For $i\in[n]$, write
$X_{<i}=(X_0,\ldots,X_{i-1})$ and $Y_{<i}=(Y_0,\ldots,Y_{i-1})$.
For $n\geq2$, applying the same fact inside the expectation separates
the next coordinate:
\begin{align*}
&\kldiv{(X_0,\ldots,X_n)}{(Y_0,\ldots,Y_n)}\\
&\quad=\kldiv{X_0}{Y_0}
+\E_{x_0\sim X_0}
\kldiv{X_1\mid_{X_0=x_0}}{Y_1\mid_{Y_0=x_0}}\\
&\qquad+\E_{(x_0,x_1)\sim(X_0,X_1)}
\kldiv{(X_2,\ldots,X_n)\mid_{X_0=x_0,\,X_1=x_1}}
       {(Y_2,\ldots,Y_n)\mid_{Y_0=x_0,\,Y_1=x_1}}\\
&\quad=\sum_{i=1}^n\E_{x_{<i}\sim X_{<i}}
\kldiv{X_i\mid_{X_{<i}=x_{<i}}}{Y_i\mid_{Y_{<i}=x_{<i}}}.
\end{align*}
The Markov property gives
\begin{align*}
X_i\mid_{X_{<i}=x_{<i}}
&\equiv X_i\mid_{X_{i-1}=x_{i-1}},\\
Y_i\mid_{Y_{<i}=x_{<i}}
&\equiv Y_i\mid_{Y_{i-1}=x_{i-1}}.
\end{align*}
Consequently, tuples $x_{<i}$ with the same last value $x_{i-1}=x$
give the same KL divergence. Their weights in the expectation add up to
$\Pr[X_{i-1}=x]$:
\begin{align*}
&\E_{x_{<i}\sim X_{<i}}
\kldiv{X_i\mid_{X_{<i}=x_{<i}}}{Y_i\mid_{Y_{<i}=x_{<i}}}\\
&\quad=\sum_x
\left(\sum_{\substack{x_{<i}:\,x_{i-1}=x}}
\Pr[X_{<i}=x_{<i}]\right)
\kldiv{X_i\mid_{X_{i-1}=x}}{Y_i\mid_{Y_{i-1}=x}}\\
&\quad=\E_{x\sim X_{i-1}}
\kldiv{X_i\mid_{X_{i-1}=x}}{Y_i\mid_{Y_{i-1}=x}}.
\end{align*}
Substituting these equalities into the expansion proves
\eqref{eq:kl-chain-rule-markov}.
\end{proof}

The KL inequality below follows from the log-sum
inequality~\cite[Theorem~2.7.1]{cover-thomas}.
\begin{fact}[Data processing inequality]\label{fact:DPI}
Let $P_X,Q_X$ be distributions on a finite set $\mathcal X$.
Let $P_Y,Q_Y$ be obtained by applying the same randomized map from
$\mathcal X$ to a finite set $\mathcal Y$. Then
$\kldiv{P_Y}{Q_Y}\leq\kldiv{P_X}{Q_X}$.
\end{fact}

\subsection{Laplace Estimator}

The Laplace estimator allows us to smooth over the cases in which the KL-divergence may be infinity, i.e. it gives every sample non-zero probability mass to prevent this occurrence. 
In this preliminary section, we recall its definition, basic properties, and the Boolean circuit complexity of the Laplace estimator.
In particular, we recall results that imply that we can efficiently batch-sample the Laplace estimator. 

\begin{definition}[Laplace estimator; see e.g.~\cite{canonne2023concentration}]\label{def:laplace-estimator}
    Let $P$ be a distribution over a finite domain $\Omega$ of size $|\Omega|=K$. For $N>0$, let $\bar{x} = x^{(1)}, \dots x^{(N)}$ be i.i.d. samples according to $P$. The Laplace estimator is defined (w.r.t. $\bar{x}$) for any $\omega \in \Omega$ as 
  \begin{equation}\label{eqn:laplace-estimator}
      \hat{P}_N(\omega) = \frac{1}{N + K} \left(1 + \sum_{j=1}^N \mathds{1}(x^{(j)}=\omega)\right) 
  \end{equation}
\end{definition}
\noindent

The following claim extends the standard expected KL bound for the
Laplace estimator (the case $Q=P$; see~\citealp[Proposition~4]{mourtada-gaiffas})
to an arbitrary target distribution $Q$.
\begin{claim}\label{claim:bound-expected-kl-laplace}
Let $P,Q$ be distributions over a finite domain $\Omega$ of size $K\geq1$,
and let $N\geq1$ be an integer. Let $\hat P_N$ be the Laplace estimate
from $N$ independent samples of $P$. Then
\[
\E\kldiv{Q}{\hat P_N}
\leq\kldiv{Q}{P}+\log\frac{N+K}{N+1}
\leq\kldiv{Q}{P}+\frac{(K-1)\log e}{N+1},
\]
where the expectation is over the samples and logarithms have base two.
\end{claim}
\begin{proof}
If $Q$ is not absolutely continuous with respect to $P$, the right-hand
side is infinite and the claim holds. Otherwise, fix $x\in\supp(Q)$
and put $p=P(x)>0$. Its sample count $V_x$ has distribution
$\operatorname{Bin}(N,p)$, and $\hat P_N(x)=(V_x+1)/(N+K)$. Hence
\begin{align*}
\E\left[\frac{P(x)}{\hat P_N(x)}\right]
&=p(N+K)\E\left[\frac1{V_x+1}\right]\\
&=\frac{N+K}{N+1}\bigl(1-(1-p)^{N+1}\bigr)
\leq\frac{N+K}{N+1}.
\end{align*}
The second equality follows by expanding the binomial expectation and
using $\binom Nv/(v+1)=\binom{N+1}{v+1}/(N+1)$.
Expanding KL divergence and applying Jensen's inequality now gives
\begin{align*}
\E\kldiv{Q}{\hat P_N}
&=\kldiv{Q}{P}
  +\sum_{x\in\supp(Q)}Q(x)\E\log\frac{P(x)}{\hat P_N(x)}\\
&\leq\kldiv{Q}{P}
  +\sum_{x\in\supp(Q)}Q(x)\log\E\left[\frac{P(x)}{\hat P_N(x)}\right]\\
&\leq\kldiv{Q}{P}+\log\frac{N+K}{N+1}.
\end{align*}
Finally, $\log(1+y)\leq y\log e$ for $y\geq0$ gives the second bound.
\end{proof}

\begin{lemma}[Conditional KL bound from joint Laplace estimation]
\label{lem:small-KL-conditional-distr-from-joint-laplace}
    Let $\mathcal{A}$ and $\mathcal{B}$ be finite domains, and $K=\lvert \mathcal{A} \rvert \cdot \lvert \mathcal{B} \rvert$. Let $P,Q$ be two distributions over $\mathcal{A} \times \mathcal{B}$. Let $(Z_t, T_t)_{t=1}^M$ be i.i.d. samples from $Q$, and let $\widehat{Q}$ and $\widehat Q_{\mathcal{B} \mid \mathcal{A} }(\cdot \mid x)$ denote the corresponding joint and conditional (given $x \in \mathcal{A}$) Laplace estimates respectively.
    Then,
    \[
    \E_{(Z_t, T_t)^M_{t=1} \sim Q^M} \, \E_{x \sim P_{\mathcal{A}}}\kldiv{P_{\mathcal{B}\mid \mathcal{A}}(\cdot \mid x)}{\widehat{Q}_{\mathcal{B}\mid \mathcal{A}}(\cdot \mid x)} \leq \kldiv{P}{Q} + \frac{K-1}{M+1}\log e.
    \]
\end{lemma}
\begin{proof}
    For any realization of the sample pairs $(Z_t,T_t)^M_{t=1}$, by the chain rule of KL divergence,
    \[
     \E_{x \sim P_{\mathcal{A}}}\kldiv{P_{\mathcal{B}\mid \mathcal{A}}(\cdot \mid x)}{\widehat{Q}_{\mathcal{B}\mid \mathcal{A}}(\cdot \mid x)} \leq \kldiv{P}{\widehat Q}.
    \]
    Moreover, by \Cref{claim:bound-expected-kl-laplace},
    \[
    \E_{(Z_t, T_t)^M_{t=1} \sim Q^M} \kldiv{P}{\widehat Q} \leq \kldiv{P}{Q} + \frac{K-1}{M+1}\log e.
    \]
    Putting the two bounds together gives the promised result.
\end{proof}

We now give an analysis of the circuit complexity of sampling the Laplace estimator.
Recall that given $N$ samples from an alphabet of size $K$, the Laplace estimator outputs one of $K$ output symbols with probability exactly given by one plus the number of its occurrences in the $N$ samples divided by $N+K$.

A simple circuit with unbounded-fan-in OR gates implements this estimator with $N + K$ input wires that are all set to zero except for a single input wire that is set to $1$.
This circuit has one unbounded fan-in OR gate for each of the $K$ output symbols. 
Enumerate the $K$ OR gates.
For $i \in [K]$, connect the $i$th input wire as an input to the $i$th OR gate.
Then, letting each of the remaining $N$ input wires represent the samples, connect these to their corresponding symbol. 
This circuit exactly implements the Laplace estimator.

However, we consider circuit complexity in terms of fan-in two and in terms of input wires that are uniformly distributed on $\{0, 1\}$.
This results in the following fact. 
\begin{fact}
    Let $P$ be any distribution over a finite domain of size $K$. For any constant $c \geq 0$, for any $N = O(K^c)$ samples sampled from $P$, there exists a Boolean circuit $C$ depending only on the $N$ sample values and the domain size $K$ with $
\tilde{O}(N + K)$ two-input gates such that $C$ on $\lambda=\lceil(\log_2 N)^2\rceil
+\lceil\log_2(N+K)\rceil+1$
independent uniform random input wires samples a distribution that is $2^{-\log^2 N}$ KL-distance within the Laplace estimator define above. 
\end{fact}
\begin{proof}
    Fix any $N$ samples $x_1, \ldots x_N \in [K]$.
    
    Given $\lambda \coloneqq \lceil \log^2 N \rceil + \lceil \log_2 (N+ K) \rceil + 1$ input wires uniform randomly distributed, let $U$ be uniformly distributed over $[0,\ldots, 2^\lambda - 1]$,  compute $J \coloneqq 1 + \lfloor (N+K) \cdot U \cdot 2^{-\lambda} \rfloor \in [N+K]$.
    If $J \leq K$ output the $J$-th symbol. 
    If $J > K$, output the symbol $x_{J-K}$.

    The total Boolean circuit complexity is $\tilde{O}(N+K)$.
    Multiplying $U$ with $N+K$ takes $O(\lambda \cdot \log(N+K))$ gates, dividing by $2^{\lambda}$ is throwing away the lower $\lambda$ bits using zero gates, and adding $1$ takes $\log (N+K)$ gates.
    Encode $J$ by the $\lceil\log_2(N+K)\rceil$-bit representation of $J-1$. First convert this encoding into $N+K$ wires, exactly one of which is one, using $O(N+K)$ gates. Combine the wires corresponding to each symbol using $N$ two-input OR gates in total, and encode the resulting $K$ symbol-indicator wires in binary using $O(K)$ additional gates. Thus this conversion costs $O(N+K)$ gates.
   
    We now bound the KL-divergence.
    Let $\widetilde P(\sigma)=\Pr[C(U)=\sigma]$ for $\sigma \in [K]$.
    Let $\widehat P$
    be the Laplace estimate of the samples $x_1, \ldots x_N$. 
    Each index $j\in[N+K]$ satisfies
    \[
    \Pr[J=j]\leq\frac{1}{N+K}+\frac{1}{2^\lambda}.
    \]
    Summing over the $1+ \abs{\{i \in [N]:x_i= \sigma\}}$ indices that output
    symbol $\sigma$, we obtain
    \[
    \widetilde P(\sigma)
    \leq
    \left(1+\frac{N+K}{2^\lambda}\right)\widehat P(\sigma)
    \leq
    (1+2^{-\log^2 N-1})\widehat P(\sigma).
    \]
    Therefore, opening the definition of KL divergence,
    \begin{align*}
    D_{\mathrm{KL}}(\widetilde P\Vert\widehat P)
    &=\sum_{\sigma\in[K]}\widetilde P(\sigma)
      \log_2\frac{\widetilde P(\sigma)}{\widehat P(\sigma)}\\
    &\leq\sum_{\sigma\in[K]}\widetilde P(\sigma)
      \log_2(1+2^{-\log^2 N-1})\\
    &\leq\sum_{\sigma\in[K]}\widetilde P(\sigma)\,2^{-\log^2 N}\\
    &=2^{-\log^2 N},
    \end{align*}
    where we used $\log_2(1+t)\leq t\log_2 e\leq2t$ for any $t > 0$.
\end{proof}

If we naively sample the Laplace estimator $N$ times, the resulting circuit complexity would blow-up by $N$, resulting in a complexity of $\tilde{O}(N(N+K))$.
This complexity is primarily due to a table look-up, i.e. the above algorithm looks up one of $K$ symbols in a table of size $N+K$.
However, it has long been known how to batch table look-ups to remove this overhead.
We now recall the sorting-network bound used for this construction.

A sorting network is a fixed sequence of compare-and-swap operations:
each operation takes two binary integers and returns their minimum and
maximum. The locations compared do not depend on the input values.
Mazor and Pass~\cite[Lemma~10]{ITCS:MazPas24} recall a sorting-circuit
bound of $O(N\cdot\operatorname{poly}(\log N,w))$ for $N$ integers of
$w$ bits each. We use the following explicit bound from Batcher's
construction to keep track of the logarithmic factors.

\begin{theorem}[Sorting networks~\cite{batcher1968sorting}]
\label{thm:sorting-networks}
For every pair of integers $N,w\geq1$, there is a sorting network with
$O(N\log^2(2N))$ compare-and-swap operations that sorts any $N$
$w$-bit integers in nondecreasing order. Each operation can be
implemented using $O(w)$ two-input Boolean gates. Consequently, there
is a deterministic Boolean circuit $\mathsf{Sort}_{N,w}$ of size
$O \left(Nw\log^2(2N)\right)$
that outputs a sorted permutation of its $N$ inputs. The circuit
construction is deterministic and depends only on $N,w$.
\end{theorem}

\begin{corollary}[Batched table lookup]
\label{cor:batched-table-lookup}
Let $M,K,L\geq1$ be integers with $2\leq K\leq M$.
There exists a Boolean circuit depending only on $M,K,L$ that,
on binary-encoded inputs $a=(a_1,\ldots,a_M)\in[K]^M$ and
$(J_1,\ldots,J_L)\in[M]^L$, outputs
$(a_{J_1},\ldots,a_{J_L})$, using
$O((M+L)\log^3(2(M+L)))=\widetilde O(M+L)$ two-input gates.
The circuit construction is deterministic.
\end{corollary}

\begin{proof}
Set $R=M+L$. A record is a tuple $(d,\tau,q,v)$ whose coordinates
are a table address $d$, a type bit $\tau$ ($0$ for a table entry,
$1$ for a query), a query index $q$, and a value $v$.
Unused coordinates are set to zero. Form the $R$ records
\[
(j,0,0,a_j)\quad(j\in[M]),
\qquad
(J_q,1,q,0)\quad(q\in[L]).
\]
For $h=\lceil\log(R+1)\rceil$, represent $(d,\tau,q,v)$ by the
$4h$-bit integer
\[
d\,2^{3h}+\tau\,2^{2h}+q\,2^h+v.
\]
Since $d,q,v\leq R<2^h$, numerical order on these encodings agrees
with lexicographic order on the tuples, and $4h=O(\log(2R))$.

Apply \Cref{thm:sorting-networks} to sort these records
lexicographically. Thus, for each $q\in[L]$, the query record
$(J_q,1,q,0)$ occurs after $(J_q,0,0,a_{J_q})$, with no other
table, i.e. type bit $0$, between them.
Write the sorted records as $(d_r,\tau_r,q_r,v_r)_{r=1}^R$,
and compute
\[
c_0=0,\qquad
c_r=
\begin{cases}
v_r,&\tau_r=0,\\
c_{r-1},&\tau_r=1.
\end{cases}
\]
For a query record with $q_r=q$, we have $c_r=a_{J_q}$.
This computation uses $O(R\log K)$ gates.

Finally, sort the triples $(1-\tau_r,q_r,c_r)$ lexicographically,
using the integer encodings $(1-\tau_r)2^{2h}+q_r2^h+c_r$.
The first $L$ triples correspond to queries in increasing order
of $q$, so their third coordinates give $(a_{J_1},\ldots,a_{J_L})$.
Each of the two sorting networks uses $O(R\log^3(2R))$ gates,
giving the claimed size bound.
All operations are fixed by $M,K,L$, so the circuit construction
is deterministic.
\end{proof}

\begin{corollary}[Batched sampling from a Laplace estimate]
\label{cor:batched-laplace-sampling}
Let $N,K,\lambda\geq1$. There exists a Boolean circuit depending only
on $N,K,\lambda$, with $\widetilde O(\lambda(N+K))$ two-input gates,
that takes binary-encoded samples $x_1,\ldots,x_N\in[K]$ as inputs
and uses $N\lambda$ independent uniform random bits. For every fixed
input, its $N$ outputs are independent with a common distribution
$\widetilde P$ satisfying
\[
    D_{\mathrm{KL}}\!\left(
        \widetilde P^{\otimes N}
        \,\middle\|\,
        \widehat P^{\otimes N}
    \right)
    \leq 2N(N+K)\,2^{-\lambda}. 
\]
Here, $\widehat P$ is the Laplace estimate of the input samples,
and $\widetilde O$ hides polylogarithmic factors in $N$ and $K$.
\end{corollary}

\begin{proof}
For $K=1$, output $N$ copies of the sole symbol.
Henceforth assume $K\geq2$.
Write $T=N+K$, and form the table $a=(1,\ldots,K,x_1,\ldots,x_N)\in[K]^T$.
For each $i\in[N]$, use a separate block of $\lambda$
random bits to generate
$U_i$ uniformly in $\{0,\ldots,2^\lambda-1\}$, and compute
$J_i=1+\left\lfloor\frac{TU_i}{2^\lambda}\right\rfloor$.
Computing all indices requires
$O(N\lambda\log(2T))$ two-input gates.
By Corollary~\ref{cor:batched-table-lookup}, computing
$(a_{J_1},\ldots,a_{J_N})$ requires
$O(T\log^3(2T))$ additional gates.
The total size is therefore
$\widetilde O(\lambda T)$.

For every fixed input, the outputs are independent, since each equals
$a_{J_i}$ and the indices are independent.
Moreover, for every $j\in[T]$, $\Pr[J_i=j]\leq \frac1T+2^{-\lambda}$.
Summing over the table positions containing $\sigma$ gives $\widetilde P(\sigma)
    \leq (1+T2^{-\lambda})\widehat P(\sigma)$.
Consequently,
\begin{align*}
D_{\mathrm{KL}}\!\left(
    \widetilde P^{\otimes N}
    \,\middle\|\,
    \widehat P^{\otimes N}
\right)
&=N D_{\mathrm{KL}}(\widetilde P\|\widehat P)\\
&\leq N\log_2(1+T2^{-\lambda})\\
&\leq 2NT\,2^{-\lambda},
\end{align*}
using the fact that $\log_2(1+t)\leq 2t$ for $t > 0$.
\end{proof}

The same construction allows an input bit for each sample to
specify whether it is used. We record this extension with the KL
bound needed below.

\begin{corollary}[Batched Laplace sampling from selected samples]
\label{cor:batched-laplace-sampling-inputs}
Let $N,K,\lambda\geq1$, $T=N+K$, and $\eta=T2^{-\lambda} < 1$.
There is a Boolean circuit depending only on $N,K,\lambda$, of size
\[
O\left(T\log^3(2T)+N\lambda\log(2T)\right)
=\widetilde O(\lambda T),
\]
that takes binary-encoded samples $x_1,\ldots,x_N\in[K]$ and
selection bits $b_1,\ldots,b_N\in\zo$ as inputs, where $b_t=1$
means that $x_t$ is included in the estimate.
Set $H=\sum_{t=1}^N b_t$, and let $\widehat P$ be the Laplace estimate
of the selected samples:
\[
\widehat P(\sigma)
=\frac{1+\sum_{t=1}^N b_t\mathbf1[x_t=\sigma]}{H+K}
\quad(\sigma\in[K]).
\]
Thus $\widehat P$ is uniform when $H=0$.

Using $N\lambda$ independent
uniform random bits, the Boolean circuit outputs $N$ symbols that, conditional on these
inputs, are independent with a common distribution $\widetilde P$.
This distribution depends only on the selected samples, in their
original order.
Then, for every
distribution $Q$ over $[K]$,
\[
\kldiv{Q}{\widetilde P}
\leq\kldiv{Q}{\widehat P}+\log_2\frac1{1-\eta}.
\]
The circuit construction is deterministic, and $\widetilde O$ hides
polylogarithmic factors in $N$ and $K$.
\end{corollary}

\begin{proof}
The case $K=1$ is immediate. Assume $K\geq2$ and compute
$H=\sum_{t=1}^N b_t$.
For $h=\lceil\log(2T)\rceil$, represent each triple
$(1-b_t,t-1,x_t-1)$ by the $(2h+1)$-bit integer
\[
(1-b_t)2^{2h}+(t-1)2^h+(x_t-1),
\]
and sort these integers. This sorts the triples lexicographically.
The first coordinate places the selected samples first.
The second coordinate preserves their original order, and the third
coordinate is the sample encoded as $x_t-1$.
Let $y_1,\ldots,y_N$ be the corresponding samples in sorted
order. Then $y_1,\ldots,y_H$ are exactly the selected samples,
in their original order.
By \Cref{thm:sorting-networks}, sorting costs
$O(N\log^2(2N)\log(2T))$ gates; computing $H$ costs
$O(N\log(2N))$ gates.

Use the construction of \Cref{cor:batched-laplace-sampling} with
the table $a=(1,\ldots,K,y_1,\ldots,y_N)$, but compute
\[
J_i=1+\left\lfloor\frac{(H+K)U_i}{2^\lambda}\right\rfloor
\qquad(i\in[N]).
\]
Here the $U_i$ are independent and uniform in
$\{0,\ldots,2^\lambda-1\}$.
The indices lie in $[H+K]$, so the lookups are restricted to
$(1,\ldots,K,y_1,\ldots,y_H)$; this also handles $H=0$.
Computing the indices costs $O(N\lambda\log(2T))$ gates,
and the lookups cost $O(T\log^3(2T))$ gates.
The sorting and counting costs are absorbed in this bound.
The outputs remain independent conditional on the inputs, and their
common distribution depends only on the selected samples
in their original order.

For every $j\in[H+K]$, the lower rounding bound gives
\[
\Pr[J_i=j]\geq\frac1{H+K}-2^{-\lambda}
\geq\frac{1-\eta}{H+K}.
\]
Summing over the positions containing $\sigma$ gives
$\widetilde P(\sigma)\geq(1-\eta)\widehat P(\sigma)$. For $\eta<1$,
\[
\kldiv{Q}{\widetilde P}
=\kldiv{Q}{\widehat P}
+\sum_{\sigma\in[K]}Q(\sigma)
\log_2\frac{\widehat P(\sigma)}{\widetilde P(\sigma)}
\leq\kldiv{Q}{\widehat P}+\log_2\frac1{1-\eta}.
\]
\end{proof}

\subsection{One-Way Functions}
\label{subsec:one-way-functions}

We use the length-preserving form of one-way functions, with security
against nonuniform circuits.

\begin{definition}[One-way functions against circuits]
\label{def:one-way-functions-circuits}
Let $f=(f_\lambda)_{\lambda\geq1}$ be a function family with
$f_\lambda:\{0,1\}^\lambda\to\{0,1\}^\lambda$, where $f_\lambda(x)$
is computable from $(1^\lambda,x)$ in deterministic polynomial time.
For functions $T:\bN\to[1,\infty)$ and $\varepsilon:\bN\to[0,1]$, we say
that $f$ is $(T,\varepsilon)$-one-way against circuits if, for all
sufficiently large $\lambda$, every randomized circuit $A_\lambda$ of
size at most $T(\lambda)$ satisfies
\[
\Pr_{X\gets U_\lambda,R}\left[
A_\lambda(f_\lambda(X);R)\in f_\lambda^{-1}(f_\lambda(X))
\right]\leq\varepsilon(\lambda).
\]
Here $R$ denotes independent uniform random coins for $A_\lambda$.
The circuit may depend arbitrarily on $\lambda$, and success means
finding any preimage; an output $\bot$ counts as failure.

We call $f$ a \emph{one-way function against polynomial-size circuits}
if it is $(\lambda^c,\lambda^{-c})$-one-way for every constant $c>0$.
For a size bound $T$, we call $f$ \emph{$T$-secure} if it is
$(T,1/T)$-one-way.
\end{definition}

\Cref{thm:insertional-vs-sequential} uses one-wayness against
polynomial-size circuits. \Cref{thm:sm-insertional-separation} assumes
the stronger subexponential security $T(\lambda)=2^{\gamma\lambda^\alpha}$
for constants $\gamma>0$ and $1/2<\alpha<1$.

\subsection{Pseudoentropy}

\begin{definition}[Pseudoentropy, nonuniform setting~\cite{vadhan-zheng}]
\label{def:pseudoentropy-nonuniform}
Let $X$ be a random variable over a finite set. We say that $X$ has
$(T,\varepsilon)$ nonuniform pseudoentropy at least $k$ if there exists a
random variable $Y$ over the same set with
\begin{enumerate}
    \item $H(Y)\geq k$ and
    \item $X$ and $Y$ are $(T,\varepsilon)$ nonuniformly indistinguishable.
\end{enumerate}
Here indistinguishability
means that every Boolean circuit of size at most $T$ has distinguishing
advantage at most $\varepsilon$ in absolute value.

If $X=X(n)$, we say that $X$ has nonuniform pseudoentropy at least $k=k(n)$
if, for every constant $c>0$, it has $(n^c,n^{-c})$ nonuniform pseudoentropy
at least $k(n)-n^{-c}$ for all sufficiently large $n$.
\end{definition}

\begin{definition}[Conditional pseudoentropy, nonuniform setting~\cite{vadhan-zheng}]
\label{def:conditional-pseudoentropy-nonuniform}
Let $(X,B)$ take values in the finite set $\mathcal X\times\mathcal B$.
We say that $B$ has
$(T,\varepsilon)$ conditional pseudoentropy at least $k$ given $X$ if there
exists a random variable $Z$ over $\mathcal B$, jointly distributed with $X$,
such that
\begin{enumerate}
    \item $H(Z\mid X)\geq k$ and
    \item $(X,B)$ and $(X,Z)$ are $(T,\varepsilon)$ nonuniformly indistinguishable.
\end{enumerate}
The conditional pseudoentropy is the supremum of such $k$. The
$(T,\varepsilon)$ conditional pseudoentropy gap of $B$ given $X$ is this
supremum minus $H(B\mid X)$.
The asymptotic definition uses the same quantifiers and entropy slack as
\Cref{def:pseudoentropy-nonuniform}.
\end{definition}

The maximum exists: the conditional distributions of $Z$ form a compact set, the
indistinguishability constraints are closed, and conditional entropy is
continuous. Taking $Z=B$ shows that the gap is nonnegative.

\begin{definition}[{Next-block pseudoentropy~\cite[Definition~2.12]{vadhan-zheng}}]
\label{def:nbp}
Let $n,\ell\geq1$, let $Y=(Y_1,\ldots,Y_n)$ take values in
$(\{0,1\}^\ell)^n$, and let $I$ be uniform on $[n]$, independently of $Y$.
We say that $Y$ has $(t,\epsilon)$ next-block pseudoentropy at least $k$
if there exists a random variable $Z$ over $\{0,1\}^\ell$, jointly
distributed with $(I,Y)$, such that
\begin{enumerate}
    \item $H(Z\mid Y_{<I})\geq k/n$ and
    \item $(Y_{<I},Y_I)$ and $(Y_{<I},Z)$ are $(t,\epsilon)$ nonuniformly indistinguishable.
\end{enumerate}
For $\ell=1$, we also call this next-bit pseudoentropy.
\end{definition}

For families indexed by $n$, next-block pseudoentropy at least $k(n)$ means
that, for every $c>0$ and all sufficiently large $n$, the
$(n^c,n^{-c})$ next-block pseudoentropy is at least $k(n)-n^{-c}$.
This is the nonuniform asymptotic convention; polynomial rescaling absorbs
the number of blocks when comparing it with the conditional definition.

\begin{definition}[KL-hardness for conditional sampling~\cite{vadhan-zheng}]
Let $(X,B)$ take values in $\{0,1\}^m\times[q]$. We say that $B$ is
$(t,\delta)$ KL-hard to sample given $X$ if every randomized circuit
$S:\{0,1\}^m\to[q]$ of size at most $t$ satisfies
$\kldiv{(X,B)}{(X,S(X))}>\delta$.
\end{definition}

We use the following two consequences of the Vadhan--Zheng characterization,
stated directly in terms of conditional samplers. In the first part, the
distinguishing error $\epsilon$ and the additive KL-divergence loss $\eta$
are separate parameters.

\begin{theorem}[Characterizing conditional pseudoentropy, nonuniform setting]
\label{thm:formal-characterizing-pseudoentropy-nonuniform}
There exist a universal constant $c_1\in(0,1]$ and a polynomial $p_0$ with
nonnegative coefficients and constant term at least one such that the following
holds. Let $(X,B)$ take
values in $\{0,1\}^m\times[q]$, with $m\geq0$ and $q\geq2$, and let
$t\geq1$, $\delta,g\geq0$, and $0<\epsilon,\eta,\sigma<1$.
\begin{enumerate}
    \item If $B$ has $(t,\epsilon)$ conditional pseudoentropy gap at least
    $\delta$ given $X$, then every randomized circuit
    $S:\{0,1\}^m\to[q]$ of size at most
    \[
    \frac{\min\{t^{c_1},\,\sigma/\epsilon\}}
         {p_0(m,q,1/\eta,\log(1/\sigma))}
    \]
    satisfies $\kldiv{(X,B)}{(X,S(X))}>\delta-\sigma-\eta$.
    \item If the $(t,\eta)$ conditional pseudoentropy gap of $B$ given
    $X$ is at most $g$, then there exists a randomized circuit
    $S:\{0,1\}^m\to[q]$ of size at most $p_0(t,m,q,1/\eta)$ such that
    \[\kldiv{(X,B)}{(X,S(X))}\leq g+2\eta.\]
\end{enumerate}
\end{theorem}
\begin{proof}
The first part combines~\cite[Theorem~3.8 and Lemma~3.6]{vadhan-zheng},
with respective KL-divergence losses $\sigma$ and $\eta$. Choosing $c_1$
sufficiently small and $p_0$ sufficiently large gives the stated circuit
sizes. For $m=0$, add a fixed conditioning bit; for $\delta=0$, the first
part follows from nonnegativity of KL-divergence.

For the second part, suppose that every sampler of size at most
$p_0(t,m,q,1/\eta)$ has KL-divergence greater than $g+2\eta$. For sufficiently
large $p_0$, the same results give a $(t,\eta)$ conditional pseudoentropy
gap at least $g+\eta$, contradicting its upper bound $g$. One polynomial
$p_0$ can be chosen large enough for both parts, absorbing constant shifts
of its arguments.
\end{proof}

The running time of the resulting sampler is polynomial in the inverse
accuracy. Consequently, the finite-parameter statement supplies a family
of polynomial-size samplers for inverse-polynomial accuracies.

\section{Sequential Sampling Algorithms}

\begin{definition}[$(\ell, s, B)$-Sequential Sampling Algorithm]
    We define a sequential sampling algorithm of size $B$ to be a sequence of randomized boolean circuits $C(1^{\outLen}) = (C_1, C_2, \ldots, C_\outLen)$ for an output length $\outLen \in \bN$ such that all circuits $C_i$ are of size at most $B$.

    We use $C(1^{\outLen})$ to denote the distribution of $(\outSym_{1},\dots,\outSym_{\outLen})$ sampled iteratively as follows: for $i=1$, sample $(\outSym_{1}, \state_{1}) \gets C_{1}(1^{\outLen})$ and for $i>1$ sample $(\outSym_{i}, \state_{i}) \gets C_{i}(1^{\outLen}, \state_{i-1}, \outSym_{1},\dots,\outSym_{i-1})$. We use $\outSymLen$ and $\stateLen$ to denote the lengths of the $x_{i}$'s and $\state_{i}$'s respectively.
\end{definition}

All invocations use fresh independent random coins. The input $1^\outLen$
is fixed for each circuit family and is suppressed when convenient.

\begin{remark}
    When $s=0$, we refer to an $(\ell, 0, B)$-\textit{sequential sampler} simply as a \textit{stateless $(\ell, B)$-\textit{sequential sampler}.}
\end{remark}

\subsection{Characterization of Sequential Sampling}
\label{sec:sequential-characterization}

We apply \Cref{thm:formal-characterizing-pseudoentropy-nonuniform} with
the visible prefix as the conditioning variable and the next block as the
variable to be sampled. Choosing the block uniformly gives a characterization
of stateless sequential sampling. The dependence on the alphabet size
$2^\ell$ is polynomial, contributing a polynomial factor in $n$ when
$\ell=O(\log n)$.

\begin{theorem}[Characterization of stateless sequential sampling]
\label{thm:sequential-characterization}
\label{thm:stateless-case-larger-alphabet}
\label{thm:from-small-NBPE-gap-to-stateless-seq-samplability}
There exist a universal constant $c_0\in(0,1]$ and a polynomial $p$ with
nonnegative coefficients and constant term at least one such that the following
holds. Let $n,\ell\geq1$, let $Y=(Y_1,\ldots,Y_n)$ take
values in $(\{0,1\}^\ell)^n$, and let $t\geq1$, $\delta\geq0$, and
$0<\epsilon,\eta,\sigma<1$.
\begin{enumerate}
    \item If $Y$ has $(t,\epsilon)$ next-block pseudoentropy at least
    $H(Y)+\delta$, then, for every $B\geq1$ satisfying
    \[
    B\leq\frac{\min\{t^{c_0},\,\sigma/(n\epsilon)\}}
                   {p(n,2^\ell,1/\eta,\log(n/\sigma))},
    \]
    every stateless $(\ell,B)$-sequential sampler $C$ satisfies
    $\kldiv{Y}{C(1^n)}>\delta-\sigma-n\eta$.
    \item If $\delta\geq\epsilon$ and the $(t,\epsilon)$ next-block
    pseudoentropy of $Y$ is at most $H(Y)+\delta-\epsilon$, then there
    exists a stateless $(\ell,B)$-sequential sampler $C$ with
    \[
    B\leq p(t,n,2^\ell,1/\epsilon),
    \qquad \kldiv{Y}{C(1^n)}\leq\delta.
    \]
\end{enumerate}
\end{theorem}

We prove the two directions separately. Throughout, $I$ is uniform on $[n]$
and independent of $Y$, and $X=(I,Y_{<I})$. Write $m=O(n\ell)$ for
the input length of $X$ and $q=2^\ell$ for the alphabet size.

\begin{lemma}[Next-block pseudoentropy implies conditional sampling hardness]
\label{lemma:seq-samplability-implies-small-NBPE-gap-v2}
Under the hypotheses and size condition of the first part of
\Cref{thm:sequential-characterization}, every stateless
$(\ell,B)$-sequential sampler $C$ satisfies
\[
\E_{i\gets[n]}\kldiv{(Y_{<i},Y_i)}{(Y_{<i},C_i(Y_{<i}))}
>\frac{\delta-\sigma}{n}-\eta.
\]
\end{lemma}
\begin{proof}
By \Cref{def:nbp}, there is a random variable $Z$ jointly distributed
with $(I,Y)$ such that $(X,Z)$ and $(X,Y_I)$ are
$(t,\epsilon)$-indistinguishable and
\[
H(Z\mid Y_{<I})
=\frac1n\sum_{i=1}^n H(Z\mid Y_{<i},I=i)
\geq\frac{H(Y)+\delta}{n}.
\]
Here $Y_{<I}$ determines $I$. Since $H(Y_I\mid X)=H(Y)/n$, the
$(t,\epsilon)$ conditional pseudoentropy gap of $Y_I$ given $X$ is at
least $\delta/n$. This is an average over the index and does not give
the same gap for each fixed $i$. Fix a stateless sampler $C$. To apply
\Cref{thm:formal-characterizing-pseudoentropy-nonuniform}, we therefore
need a sampler that, given $(I,Y_{<I})$, samples exactly
$C_I(Y_{<I})$. The loss $\eta$ in that theorem comes from approximating
a sampler's output probabilities to obtain the predictor used in its proof.

We now construct this sampler, which we call $S$. It contains one copy
of each $C_j$ and, on input $(i,y_{<i})$, selects $C_i$ to produce
the output $C_i(y_{<i})$. The size increases, but remains sufficiently
small for the first part of
\Cref{thm:formal-characterizing-pseudoentropy-nonuniform} to apply.
Namely, we implement this selection by ANDing each output bit of $C_j$
with the comparison $i=j$, and then ORing these bits over $j\in[n]$.
This sets every unselected output to zero, leaving only the output of
$C_i$.

Each $C_j$ contributes at most $B$ gates, with $O(n\ell)$ more for its
index comparison and output masking, giving $B+O(n\ell)$ gates per copy.
With $n$ copies and $O(n\ell)$ gates for the final ORs, the size of $S$
is $O(n(B+n\ell))=O(n^2qB)$.
Take $c_0=c_1$. Since $m=O(nq)$ and $p_0$ has nonnegative coefficients,
choose the fixed polynomial $p$ sufficiently large to absorb the
$O(n^2q)$ overhead and the dependence of $p_0$ on $m$.
The assumed bound on $B$ then makes the size of $S$ at most
\[
\frac{\min\{t^{c_1},\,\sigma/(n\epsilon)\}}
     {p_0(m,q,1/\eta,\log(n/\sigma))}.
\]
Therefore, the first part of
\Cref{thm:formal-characterizing-pseudoentropy-nonuniform}, with gap
$\delta/n$, loss $\sigma/n$, and the same $t$, $\epsilon$, and $\eta$,
gives
\begin{equation}
\label{eq:random-index-conditional-kl}
\begin{aligned}
\kldiv{(X,Y_I)}{(X,S(X))}
&=\frac1n\sum_{i=1}^n
  \kldiv{(Y_{<i},Y_i)}{(Y_{<i},C_i(Y_{<i}))} \\
&>\frac{\delta-\sigma}{n}-\eta.
\end{aligned}
\end{equation}
The equality conditions on the common uniform index $I$.
\end{proof}

\begin{lemma}[Small next-block pseudoentropy gap gives samplers]
\label{lemma:small-NBPE-gap-implies-seq-samplability}
Under the hypotheses of the second part of
\Cref{thm:sequential-characterization}, there is a stateless
$(\ell,B)$-sequential sampler $C$, with
$B\leq p(t,n,2^\ell,1/\epsilon)$, such that
\[
\E_{i\gets[n]}\kldiv{(Y_{<i},Y_i)}{(Y_{<i},C_i(Y_{<i}))}
\leq\frac{\delta}{n}.
\]
\end{lemma}
\begin{proof}
By assumption, the $(t,\epsilon)$ next-block pseudoentropy of $Y$ is at
most $H(Y)+\delta-\epsilon$. We use this bound in the second part of
\Cref{thm:formal-characterizing-pseudoentropy-nonuniform} to obtain a
sampler $S$ that receives $(i,y_{<i})$ and samples $Y_I$ given $X$ up to
KL-divergence $\delta/n$.
Fixing its index input to $i$ gives the circuit $C_i$: on input
$y_{<i}$, it outputs $S(i,y_{<i})$.

By \Cref{def:nbp} and $H(Y_I\mid X)=H(Y)/n$, the
$(t,\epsilon)$ conditional pseudoentropy gap of $Y_I$ given $X$ is at
most $(\delta-\epsilon)/n$. The target KL-divergence $\delta/n$ leaves
$\epsilon/n$ for the theorem's additive loss $2\eta$. We therefore set
$\eta'=\epsilon/(2n)$. Since $\eta'\leq\epsilon$, every distribution satisfying
$(t,\eta')$-indistinguishability also satisfies
$(t,\epsilon)$-indistinguishability, so the same upper bound on the gap
holds at error $\eta'$.
Since $\delta\geq\epsilon$ and $0<\eta'<1$, we can therefore
apply that theorem with $g=(\delta-\epsilon)/n$ and $\eta'$ in
place of $\eta$. It gives a randomized circuit $S$ of size at most
$p_0(t,m,q,2n/\epsilon)$ such that
\[
\kldiv{(X,Y_I)}{(X,S(X))}
\leq \frac{\delta-\epsilon}{n}+2\eta'
=\frac{\delta}{n}.
\]

For each $i\in[n]$, define $C_i(y_{<i})=S(i,y_{<i})$ by hardwiring
the index input to $i$. This does not increase the circuit size.
Using fresh independent coins at every step, the circuits $C_i$ form a
stateless sequential sampler. Since $m=O(nq)$, their size is bounded by
a fixed polynomial in $t,n,q,1/\epsilon$. Enlarge the polynomial $p$
from the preceding lemma to dominate this bound, retaining nonnegative
coefficients and constant term at least one. This is compatible with
that lemma because enlarging $p$ only strengthens its size hypothesis.
Thus $B\leq p(t,n,2^\ell,1/\epsilon)$.

Finally, the two distributions in the KL-divergence have the same
uniform index $I$. Conditioning on this index gives
\[
\E_{i\gets[n]}\kldiv{(Y_{<i},Y_i)}{(Y_{<i},C_i(Y_{<i}))}
=\kldiv{(X,Y_I)}{(X,S(X))}
\leq\frac{\delta}{n}.
\]
\end{proof}

\begin{proof}[Proof of \Cref{thm:sequential-characterization}]
Since the sampler is stateless and uses fresh coins at each step,
$\Pr[C(1^n)=y]=\prod_i\Pr[C_i(y_{<i})=y_i]$.
Together with the analogous factorization of $\Pr[Y=y]$, this gives
\begin{align*}
\label{eq:sequential-total-kl}
\kldiv{Y}{C(1^n)}
&=\sum_{i=1}^n\kldiv{(Y_{<i},Y_i)}{(Y_{<i},C_i(Y_{<i}))}
\\&=n\E_{i\gets[n]}\kldiv{(Y_{<i},Y_i)}{(Y_{<i},C_i(Y_{<i}))}.
\end{align*}
Indeed, when the divergence is finite, its definition gives
\[
\kldiv{Y}{C(1^n)}
=\sum_{y\in\supp(Y)}\Pr[Y=y]\sum_{i=1}^n
\log\frac{\Pr[Y_i=y_i\mid Y_{<i}=y_{<i}]}
               {\Pr[C_i(y_{<i})=y_i]}.
\]
For each $i$, marginalizing over $y_{>i}$ identifies its contribution as
$\kldiv{(Y_{<i},Y_i)}{(Y_{<i},C_i(Y_{<i}))}$.
If some $y\in\supp(Y)$ has zero
probability under $C$, at least one of its next-block probabilities is zero;
the corresponding summand and the total KL-divergence are both infinite.
The first conclusion now follows from
\Cref{lemma:seq-samplability-implies-small-NBPE-gap-v2}, and the second from
\Cref{lemma:small-NBPE-gap-implies-seq-samplability}.
\end{proof}

\section{Stateful Sequential Sampling}

Private state lets a sampler retain information that its visible output
has not revealed. We show how to recover the next-token distribution
from the prefix alone, at a cost essentially linear in the number of
possible states. The simulation estimates the distribution of the hidden
state conditioned on the visible prefix.

\subsection{Sequential Samplers with State}

Let $\cX$ denote the output alphabet of a block, so that
$\abs{\cX}=2^{\outSymLen}$, and let $\cS=\zo^k$ denote the state space.

\begin{definition}[Sequential sampler with $k$ bits of state]
\label{def:k-state-sequential-sampler}
A sequential sampler with $k$ bits of state and size $B$ for $\outLen$ blocks over alphabet $\cX$
is a sequence $C=(C_1,\ldots,C_\outLen)$ of randomized circuits,
each of size at most $B$, with state space $\cS=\zo^k$. The variables
$(X_1,S_1),\ldots,(X_\outLen,S_\outLen)$ are generated using independent
random coins for each circuit:
\[
(X_1,S_1)\gets C_1(1^\outLen),
\]
and for $i\ge 2$,
\[
(X_i,S_i)\gets C_i(1^\outLen,X_{<i},S_{i-1}).
\]
The sampler $C$ outputs $X=(X_1,\ldots,X_\outLen)$; the states
$S_1,\ldots,S_\outLen$ are not part of its output.
\end{definition}

To simulate $C_i$ for $i\geq2$ on a prefix $x_{<i}$ such that
$\Pr[X_{<i}=x_{<i}]>0$, it suffices to sample
$S_{i-1}$ conditioned on $X_{<i}=x_{<i}$, and then run
$C_i(1^\outLen,x_{<i},S_{i-1})$. The following lemma justifies using only
this state distribution, without reconstructing the preceding private states.

\begin{lemma}[Markov chain of sequential sampling states]
\label{lem:ssa-markov-chain}
Let $(X_1,S_1),\ldots,(X_\outLen,S_\outLen)$ be generated by a sequential
sampler with private state. For $i\in[\outLen]$, define
$A_i\coloneqq(X_{\leq i},S_i)$. Then $A_1-A_2-\cdots-A_\outLen$ is a
Markov chain.
\end{lemma}

\begin{proof}
Fix $i\in\{2,\ldots,\outLen\}$ and $a_1,\ldots,a_{i-1}$ such that
$\Pr[A_1=a_1,\ldots,A_{i-1}=a_{i-1}]>0$. Write
$a_{i-1}=(x_{<i},s_{i-1})$. For every $x_i\in\cX$ and $s_i\in\cS$,
put $a_i=(x_{\leq i},s_i)$. Since $C_i$ uses fresh random coins,
\begin{align*}
&\Pr[A_i=a_i\mid A_1=a_1,\ldots,A_{i-1}=a_{i-1}]\\
&\quad=\Pr[C_i(1^\outLen,x_{<i},s_{i-1})=(x_i,s_i)]\\
&\quad=\Pr[A_i=a_i\mid A_{i-1}=a_{i-1}].
\end{align*}
If the prefix in $a_i$ does not extend $x_{<i}$, both conditional
probabilities are zero. This proves the Markov property.
\end{proof}

\subsection{Sampling from an Empirical Distribution}

Before constructing a stateless sampler from a stateful one, we give a
sampling circuit that reduces the circuit-size overhead.
In the construction below, we run $C_i$ $m$ times to estimate the
distribution of states supplied to $C_{i+1}$.
We use the Laplace estimate (\Cref{def:laplace-estimator}) to give every
state positive probability and keep the KL error finite.
Once these samples have been obtained, a direct implementation on
a domain $[K]=\{1,\ldots,K\}$ uses $\widetilde O(m)$ gates per sample,
where the tilde hides polylogarithmic factors in $m$ and $K$.
Producing $m$ independent samples separately therefore costs
$\widetilde O(m^2)$ gates. By batching the table lookups for these draws,
we reduce this cost to $\widetilde O(m)$.
\Cref{cor:batched-laplace-sampling-inputs} gives the circuit and bounds
its additional KL error. We use it below with $N=m\geq K$, so its size
is $\widetilde O(m)$ when $\lambda=O(\log(m+K))$.

\begin{theorem}[State elimination for sequential samplers (small KL)]
\label{thm:state-elimination-KL}
Let $n\geq1$ and $k,\ell\geq0$ be integers, and let $0<\delta\leq1$. Let
$C=(C_1,\ldots,C_n)$ be a sequential sampler with $k$ bits of private
state and size at most $B$ per circuit, over alphabet
$\cX=\zo^\ell$, and let $\cS=\zo^k$.
There is a stateless sequential sampler $D=(D_1,\ldots,D_n)$
such that
$\kldiv{C(1^n)}{D(1^n)}\leq\delta
$
and every $D_i$ has Boolean circuit size at most
\[
O\left(
\frac{n^3|\cX||\cS|}{\delta}
\left[B+\log^3\!\left(\frac{n|\cX||\cS|}{\delta}\right)\right]
\right).
\]
The circuit construction is deterministic; all sampling and estimation
randomness is internal to the circuits.
\end{theorem}
The dependence on the number of states is linear up to logarithmic
factors.
The analysis estimates the joint distribution of a token and state
\emph{before} conditioning on the observed token. Applying the KL chain rule to its
Laplace estimate bounds the average error after conditioning, even when
some tokens are rare.

At step $j$, the simulation estimates the joint distribution of
$(X_j,S_j)$ conditioned on $X_{<j}=x_{<j}$, using a fixed number of samples.
Conditioning this estimate on its first component being $x_j$ gives an
estimate of the distribution of $S_j$ conditioned on $X_{\leq j}=x_{\leq j}$.
The KL chain rule bounds the expected KL error of this state estimate,
for $x_j\sim X_j\mid X_{<j}=x_{<j}$, by the KL error of the joint estimate.

\begin{proof}
Write $X=C(1^n)$. Let $D_1$ run $C_1(1^n)$
and return the first component of its output. Thus $D_1(1^n)\equiv X_1$,
proving the case $n=1$. Assume $n\geq2$ and set
\[
m=\left\lceil\frac{4n^2|\cX||\cS|}{\delta}\right\rceil.
\]
We use the circuit $\mathsf{Sample}$ from
\Cref{cor:batched-laplace-sampling-inputs} with
\[
N=m,\qquad K=|\cS|,\qquad
\lambda=\left\lceil\log\bigl(2m(m+|\cS|)\bigr)\right\rceil,
\]
identifying $[|\cS|]$ with $\cS$. For a list $L$ of at most $m$ states,
$\mathsf{Sample}(L)$ receives states $v_1,\ldots,v_m\in\cS$ and
selection bits $b_1,\ldots,b_m$ such that
$L=(v_t:b_t=1)$, in increasing order of $t$.
Write $\widehat\pi^L$ for its Laplace estimate and
$\widetilde\pi^L$ for the common distribution of the $m$ independent
outputs. Since $\eta=(m+|\cS|)2^{-\lambda}\leq1/(2m)$, the corollary gives,
for every distribution $P$ over $\cS$,
\begin{equation}
\label{eq:stateful-laplace-rounding}
\kldiv{P}{\widetilde\pi^L}
\leq\kldiv{P}{\widehat\pi^L}+\frac1m,
\end{equation}
where $\log(1/(1-1/(2m)))\leq1/m$ follows from
$(1-1/(2m))^m\geq1/2$.

In the algorithm below, a list $L_j$ formed from sampled pairs
$(Z_t,T_t)_{t=1}^m$ is represented by the states
$(T_t)_{t=1}^m$ and selection bits
$(\mathbf1[Z_t=x_j])_{t=1}^m$.
Each list preserves repetitions and sample order, and may be empty.
All lists are computed inside $D_i$; no state is passed between
the circuits $D_1,\ldots,D_n$.

For $i\geq2$, the circuit $D_i(1^n,x_{<i})$ runs the following algorithm,
using fresh independent random bits in every call:
\begin{enumerate}
\item Draw $m$ pairs from $C_1(1^n)$. Form a list $L_1$ of states, consisting of the second components
of those symbol-state pairs whose first component equals $x_1$.
\item For $j=2,\ldots,i-1$, run $\mathsf{Sample}(L_{j-1})$ to obtain
$m$ states $s_1,\ldots,s_m$. For each $t\in[m]$, run
$C_j(1^n,x_{<j},s_t)$. Form a list $L_j$ from the second components of those symbol-state
pairs whose first component equals $x_j$.
\item Run $\mathsf{Sample}(L_{i-1})$, take its first output $s$, and return the
first component of $C_i(1^n,x_{<i},s)$.
\end{enumerate}

\noindent
For
$\Pr[X_{\leq j}=x_{\leq j}]>0$, let
\[
\pi_j^{x_{\leq j}}(s)
=\Pr[S_j=s\mid X_{\leq j}=x_{\leq j}]
\]
denote the true state posterior distribution. In the following,
$[\cdot]_1$ denotes the first output component. Let
\[
\varepsilon_j
=\E_{x_{\leq j}\sim X_{\leq j}}\E_{L_j\mid x_{\leq j}}
  \kldiv{\pi_j^{x_{\leq j}}}{\widetilde\pi^{L_j}},
\]
where the expectation $\E_{L_j\mid x_{\leq j}}$ averages only over the algorithm's
random bits on input $x_{\leq j}$, independent of $X$.
By the chain rule of KL (\Cref{fact:kl-chain-rule}),
\begin{align*}
&\kldiv{C(1^n)}{D(1^n)}\\
&\quad=\sum_{i=2}^n\E_{x_{<i}\sim X_{<i}}\kldiv{X_i\mid_{X_{<i}=x_{<i}}}{D_i(1^n,x_{<i})}\\
&\quad\leq\sum_{i=2}^n\E_{x_{<i}\sim X_{<i}}\E_{L_{i-1}\mid x_{<i}}
  \kldiv{X_i\mid_{X_{<i}=x_{<i}}}
  {\bigl[C_i(1^n,x_{<i},[\mathsf{Sample}(L_{i-1})]_1)\bigr]_1},
\end{align*}
where the last inequality follows from the convexity of KL divergence, as $D_i$ constructs $L_{i-1}$ using fresh randomness.
Conditioned on $X_{<i}=x_{<i}$, by \Cref{lem:ssa-markov-chain}, $X_i$ is the first component of $C_i$ when given as input the prefix $x_{<i}$ together with a state distributed according to $\pi^{x_{<i}}_{i-1}$. Thus, for any fixed list of states $L$, by DPI (\cref{fact:DPI}),
\[
\kldiv{X_i\mid_{X_{<i}=x_{<i}}}
  {\bigl[C_i(1^n,x_{<i},[\mathsf{Sample}(L)]_1)\bigr]_1}
\leq\kldiv{\pi_{i-1}^{x_{<i}}}{\widetilde\pi^L},
\]
which together with the previous bound gives us
\begin{equation}\label{eq:kl-bound-stateful}
\kldiv{C(1^n)}{D(1^n)}\leq\sum_{j=1}^{n-1}\varepsilon_j.
\end{equation}

For $2\leq j\leq n-1$, fix $x_{<j}$ with
$\Pr[X_{<j}=x_{<j}]>0$ and a realization $L$ of the computed list $L_{j-1}$. Set
\begin{align*}
P_j(x,s)&=\Pr[X_j=x,S_j=s\mid X_{<j}=x_{<j}],\\
Q_j(x,s)&=\Pr\bigl[
C_j(1^n,x_{<j},[\mathsf{Sample}(L)]_1)=(x,s)
\bigr].
\end{align*}
By
\Cref{lem:ssa-markov-chain}, $P_j$ and $Q_j$ result from running
$C_j(1^n,x_{<j},\cdot)$ on $\pi_{j-1}^{x_{<j}}$ and
$\widetilde\pi^L$, respectively. Thus, by DPI
(\Cref{fact:DPI}),
\begin{equation}\label{eq:kl-bound-joint-stateful-dpi}
\kldiv{P_j}{Q_j}\leq\kldiv{\pi_{j-1}^{x_{<j}}}{\widetilde\pi^L}.
\end{equation}

Step~2 produces $m$ i.i.d. pairs $(Z_t,T_t)_{t=1}^m$ from $Q_j$, and forms the next list of states $L_j=(T_t : Z_t=x_j)$.
Let $\widehat Q_j$ denote the joint Laplace estimate from these pairs.
That is, letting $N_{x,s}=\sum_{t=1}^m\mathbf1[(Z_t,T_t)=(x,s)]$,
\[
\widehat Q_j(x,s)=\frac{N_{x,s}+1}{m+|\cX||\cS|}.
\]
The corresponding conditional Laplace estimate is exactly that of $L_j$:
\[
\widehat\pi^{L_j}(s)=\widehat Q_j(s\mid x_j)
=\frac{N_{x_j,s}+1}{N_{x_j}+|\cS|},
\qquad N_x=\sum_{s\in\cS}N_{x,s}.
\]
By \eqref{eq:stateful-laplace-rounding}, for every realization of $L_j$,
\[
\kldiv{\pi_j^{x_{\leq j}}}{\widetilde\pi^{L_j}}
\leq\kldiv{\pi_j^{x_{\leq j}}}{\widehat Q_j(\cdot\mid x_j)}+\frac1m.
\]
Conditional on $X_{<j}=x_{<j}$ and $L_{j-1}=L$, the sampled pairs
are independent of $X_j$.
Therefore, writing $P_{j,\cX}$ for the token marginal of $P_j$,
\begin{align*}
&\E_{(Z_t,T_t)_{t=1}^m\sim Q_j^{\otimes m}}\E_{x_j\sim P_{j,\cX}}
  \kldiv{\pi_j^{x_{\leq j}}}{\widetilde\pi^{L_j}}\\
&\quad\leq\E_{(Z_t,T_t)_{t=1}^m\sim Q_j^{\otimes m}}\E_{x_j\sim P_{j,\cX}}
  \kldiv{\pi_j^{x_{\leq j}}}{\widehat Q_j(\cdot\mid x_j)}+\frac1m\\
&\quad\leq\kldiv{P_j}{Q_j}
  +\frac{(|\cX||\cS|-1)\log e}{m+1}+\frac1m\\
&\quad\leq\kldiv{\pi_{j-1}^{x_{<j}}}{\widetilde\pi^L}
  +\frac{2|\cX||\cS|}{m},
\end{align*}
where the second-last inequality follows from \Cref{lem:small-KL-conditional-distr-from-joint-laplace} and the last inequality from \cref{eq:kl-bound-joint-stateful-dpi} and $\log e<2$.
For $j=1$, step~1 gives $P_1=Q_1=C_1(1^n)$, so the same calculation
ends at $2|\cX||\cS|/m$. For $j\geq2$, averaging over $x_{<j}\sim X_{<j}$ and the list
$L_{j-1}$ computed on that prefix gives
\begin{align*}
\varepsilon_1&\leq\frac{2|\cX||\cS|}{m},\\
\varepsilon_j&\leq\varepsilon_{j-1}+\frac{2|\cX||\cS|}{m}
  &&(2\leq j\leq n-1),\\
\varepsilon_j&\leq\frac{2j|\cX||\cS|}{m}
  &&(1\leq j\leq n-1).
\end{align*}
Therefore, by \Cref{eq:kl-bound-stateful},
\begin{align*}
\kldiv{C(1^n)}{D(1^n)}
&\leq\sum_{j=1}^{n-1}\varepsilon_j
\leq\frac{2|\cX||\cS|}{m}\sum_{j=1}^{n-1}j\\
&=\frac{n(n-1)|\cX||\cS|}{m}\leq\delta,
\end{align*}
as promised.

\paragraph{Boolean circuit size.}
Since $m\geq|\cS|$ and $\lambda=O(\log(2m))$,
\Cref{cor:batched-laplace-sampling-inputs} gives
\[
|\mathsf{Sample}|=O(m\log^3(2m)).
\]
Forming the input to $\mathsf{Sample}(L_j)$ uses $m$ copies of
$C_j$, each of size at most $B$, and $m$ comparisons with $x_j$,
each costing $O(\ell)$ gates.
The sorting and counting costs are included in $|\mathsf{Sample}|$.
There are $i-1$ such lists, $i-1$ calls to $\mathsf{Sample}$,
and one final call to $C_i$. Thus
\begin{align*}
|D_i|
&=O\left(i \cdot m \cdot \left[B+\ell+\log^3(2m)\right]\right)\\
&=O\left(
\frac{n^3|\cX||\cS|}{\delta}
\left[B+\log^3\!\left(\frac{2n|\cX||\cS|}{\delta}\right)\right]
\right),
\end{align*}
since $i\leq n$, $m=O(n^2|\cX||\cS|/\delta)$, and
$\ell=\log|\cX|=O(\log(2m))$. All calls and list representations have
fixed size, so this construction gives a stateless randomized Boolean
circuit.
\end{proof}

For polynomial-size alphabets and $k=O(\log n)$, the simulation has
polynomial size at any specified inverse-polynomial KL accuracy.
Thus every distribution generated efficiently with logarithmic private
state can be approximated efficiently without it. Together with the
known characterization by next-block pseudoentropy
(\Cref{thm:sequential-characterization}), this extends sampling
obstructions to logarithmic state; the cryptographic consequences below
apply this connection.

\subsection{Sample Complexity of Estimating the State Distribution}
\label{subsec:state-distribution-lower-bound}

The construction in \Cref{thm:state-elimination-KL} estimates the entire
conditional state distribution before sampling a state from that estimate.
We show that, when this distribution is unknown and the only access to it is
through independent samples, expected KL divergence at most $1/4$ requires
$\Omega(\abs{\cS})=\Omega(2^k)$ samples. This establishes the necessity of the
linear dependence on $\abs{\cS}$ when estimating the state distribution
from independent samples, but it does not by itself establish a matching circuit-size lower bound for \Cref{thm:state-elimination-KL}.

\begin{proposition}[Sample lower bound for estimating the state distribution]
\label{prop:state-distribution-sample-lower-bound}
Let $k\geq1$ and $N\geq0$ be integers, and let $\cS=\zo^k$. Suppose that a
randomized algorithm receives at most $N$ independent samples from an unknown
distribution $P$ over $\cS$ and outputs a vector of probabilities
$\hat\pi$ over $\cS$, with no subsequent access to the sampling oracle.
The algorithm receives no other information about $P$.
Then there is a distribution $P$ for which
\[
\E\bigl[\kldiv{P}{\hat\pi}\bigr]
\geq 1-\frac{2N}{\abs{\cS}},
\]
where the expectation is over the samples and the algorithm's internal
randomness. Consequently, if
$\E[\kldiv{P}{\hat\pi}]\leq1/4$ for every $P$, then
$N\geq3\abs{\cS}/8=3\cdot2^{k-3}$.
\end{proposition}

\begin{proof}
We partition the states into pairs and choose one state from each pair
using a hidden bit. Each sample reveals only the bit for its pair;
every pair that is never sampled contributes to the expected KL error.

Let $m=2^{k-1}$ and identify $\cS$ with $[m]\times\zo$. For each
$\theta=(\theta_1,\ldots,\theta_m)\in\zo^m$, define
\[
P_\theta(a,b)=
\begin{cases}
1/m,&b=\theta_a,\\
0,&b\neq\theta_a.
\end{cases}
\]
Choose $\Theta$ uniformly from $\zo^m$. Conditional on $\Theta$, a sample
from $P_\Theta$ is generated by choosing $A$ uniformly from $[m]$,
independently of $\Theta$, and returning $(A,\Theta_A)$.
We may generate $N$ independent samples in advance and let the algorithm
read as many as it requests, ignoring any unused samples.

Let $W$ consist of these $N$ samples and the algorithm's independent
random coins. Fix $W$, so that the output $\hat\pi$ is fixed, and let
$U\subseteq[m]$ contain the first coordinates absent from the samples.
For each $a\notin U$, the samples determine $\Theta_a$; conditional on $W$,
the bits $(\Theta_a)_{a\in U}$ remain independent and uniform.
Consequently, the distribution $\bar P=\E[P_\Theta\mid W]$ satisfies
\[
\bar P(a,b)=
\begin{cases}
1/m,&a\notin U\text{ and }b=\Theta_a,\\
0,&a\notin U\text{ and }b\neq\Theta_a,\\
1/(2m),&a\in U.
\end{cases}
\]
In particular,
\[
H(\bar P)
=\frac{m-|U|}{m}\log m+\frac{|U|}{m}\log(2m)
=\log m+\frac{|U|}{m},
\qquad H(P_\theta)=\log m.
\]
Expanding KL divergence gives
\begin{align*}
\E\bigl[\kldiv{P_\Theta}{\hat\pi}\mid W\bigr]
&=-\log m-\sum_{s\in\cS}\bar P(s)\log\hat\pi(s)\\
&=\kldiv{\bar P}{\hat\pi}+H(\bar P)-\log m
\geq\frac{|U|}{m}.
\end{align*}
At most $N$ pairs appear among the samples, so $|U|\geq m-N$. Therefore
\[
\E\bigl[\kldiv{P_\Theta}{\hat\pi}\bigr]
\geq\frac{\E|U|}{m}
\geq1-\frac Nm
=1-\frac{2N}{|\cS|}.
\]
Averaging over the uniform choice of $\Theta$ implies that some fixed
$\theta$ satisfies the same lower bound, with expectation only over its
samples and the algorithm's coins. If this expectation is at most $1/4$
for every $P_\theta$, then $1-2N/|\cS|\leq1/4$, or
$N\geq3|\cS|/8$.
\end{proof}

Every distribution $P_\theta$ used in the proof can occur as a conditional
state distribution: the first circuit can output a fixed block $x_1$,
choose $A$ uniformly from $[2^{k-1}]$, and set $S_1=(A,\theta_A)$. Then
$\pi_1^{x_1}=P_\theta$. Thus the lower bound applies even when independent
samples from $\pi_1^{x_1}=P_\theta$ are supplied directly.
The first circuit may encode $\theta$ in its description; the proposition
assumes access only to samples, rather than to that description.
\section{Insertional Sampling and Insertional Pseudoentropy}

An insertional sampler chooses both the next token and where to place it.
This abstracts the insertion framework of
\citet{DBLP:conf/icml/SternCKU19}, with one sampled insertion per step.
The current string is the only information retained between steps;
each circuit uses fresh independent coins.

\begin{definition}[Stateless Insertional Sampler]
    \label{def:statelessInsertionalSampler}
    Let $n,\ell\geq1$ and let $L$ be a distribution over $(\{0,1\}^\ell)^n$. An insertional sampler $C = (C_{1},\dots,C_{n})$ of $L$ is a sequence of randomized (possibly unbounded) circuits, where
    \begin{itemize}[noitemsep]
        \item $C(1^n) \coloneqq C_{n}(C_{n-1}(\cdots(C_1(\emptyString)))) \equiv L$, and
        \item for every $i \in [n]$ and $x \in (\{0,1\}^\ell)^{i-1}$, $C_{i}(x)$ is a distribution over strings in $(\{0,1\}^\ell)^i$ for which $x$ is a subsequence.
    \end{itemize}
    We say that $C$ has size $B$ if for all $i\in[n]$, $C_i$ is of size at most $B$.
\end{definition}

\begin{definition}[Insertional decomposition]
    \label{def:insertionalDecomposition}
    Let $n,\ell\geq1$ and let $L$ be a distribution over $(\{0,1\}^\ell)^n$. An insertional decomposition $(L_{0},\dots,L_{n})$ of $L$ is a sequence of jointly distributed random variables, where
    \begin{itemize}[noitemsep]
        \item $L_{0}$ is the fixed distribution over the empty string $\emptyString$,
        \item $L_{n} \equiv L$, and
        \item for every $i \in [n]$ and $x \in (\{0,1\}^\ell)^{i-1}$ with $\Pr[L_{i-1}=x]>0$, $L_{i} \mid L_{i-1} = x$ is a distribution over strings in $(\{0,1\}^\ell)^i$ for which $x$ is a subsequence.
    \end{itemize}
\end{definition}

For each insertional decomposition, we compare the apparent conditional
entropy of each insertion with its actual conditional entropy. The
randomness in choosing the insertion order therefore does not itself
contribute to the pseudoentropy gap.

\begin{definition}[Insertional Pseudoentropy]
    \label{def:insertionalPseudoentropy}
    Let $n,\ell\geq1$ and let $L$ be distributed over $(\{0,1\}^\ell)^n$. Let $T\geq1$, $k\geq0$, and
    $0<\varepsilon<1$; these quantities may depend on a security
    parameter $\secParam$.
    We say that $L$ has $(T,\varepsilon)$ insertional pseudoentropy at
    least $k$ if, for every insertional decomposition $(L_0,\ldots,L_n)$
    of $L$, with $I$ uniform on $[n]$ independently of the decomposition,
    there exists a random variable $Z$, jointly distributed with
    $(I,L_0,\ldots,L_n)$ and satisfying $Z\in(\{0,1\}^\ell)^i$ whenever $I=i$,
    such that
    \begin{enumerate}
        \item
        $H(Z\mid I,L_{I-1})-H(L_I\mid I,L_{I-1})
        \geq\frac{k-H(L)}{n}$,
        and
        \item $(I,L_I,L_{I-1})$ and $(I,Z,L_{I-1})$ are
        $(T,\varepsilon)$ nonuniformly indistinguishable.
    \end{enumerate}
    The insertional pseudoentropy is the supremum of the values $k$
    satisfying this definition.
\end{definition}

\begin{remark}[Comparison with next-block pseudoentropy]
\label{rem:insertional-nbpe-comparison}
Both definitions measure the increase in conditional entropy that can
remain computationally undetectable. In next-block pseudoentropy, the
entropy chain rule makes the subtraction of the actual conditional
entropy implicit. For $Y=(Y_1,\ldots,Y_n)$ with
$Y_i\in\{0,1\}^\ell$, $\ell\geq1$, and $I$ uniform on $[n]$
independently of $Y$, the length of $Y_{<I}$ determines $I$, so
\[
nH(Y_I\mid Y_{<I})
=\sum_{i=1}^nH(Y_i\mid Y_{<i})=H(Y).
\]
Consequently, the condition $H(Z\mid Y_{<I})\geq k/n$ in
\Cref{def:nbp} is equivalent to
\[
H(Z\mid Y_{<I})-H(Y_I\mid Y_{<I})
\geq\frac{k-H(Y)}{n}.
\]
This is the same entropy-gap form as in
\Cref{def:insertionalPseudoentropy}.
For an insertional decomposition, however, the entropy chain rule gives
\begin{align*}
nH(L_I\mid I,L_{I-1})
&=\sum_{i=1}^nH(L_i\mid L_{i-1})\\
&=\sum_{i=1}^n\bigl(H(L_i)-H(L_{i-1})+H(L_{i-1}\mid L_i)\bigr)\\
&=H(L)+\sum_{i=1}^nH(L_{i-1}\mid L_i).
\end{align*}
In next-block pseudoentropy, a prefix determines all prior information, whereas an in insertional pseudoentropy, a string given by $L_{i}$ does not necessarily determine its preceding string $L_{i-1}$. 
For example, for even $i$, if $L_i$ is $01$ repeated $i/2$ times, then there are $i$ distinct possibilities for $L_{i-1}$ and a random choice of $L_{i-1}$ among them would result in $H(L_{i-1} \mid L_i) = \log i$.
Therefore, the sum $\sum_{i \in [n]} H(L_{i-1} \mid L_i)$ could be as large as $\log (n!)$. Subtracting the actual conditional
entropy in \Cref{def:insertionalPseudoentropy} ensures that the gap
$k-H(L)$ does not count randomness in the choice of insertion order.
The quantification over every insertional decomposition accounts for
the sampler's freedom to choose this order.
This is implicitly the case for next-block pseudoentropy since left-to-right generation
has a single fixed prefix decomposition.
\end{remark}

The \emph{insertional pseudoentropy gap} is the insertional pseudoentropy
minus $H(L)$. Taking $Z=L_I$ shows that this gap is nonnegative.
The joint distribution of $Z$ is chosen for each decomposition before
the distinguishing circuit: the same $Z$ must fool every Boolean circuit
of size at most $T$. The guarantee also holds for randomized circuits
with independent coins, by fixing their coins and averaging. The length
of $L_{I-1}$ determines $I$, so conditioning on $(I,L_{I-1})$ is
equivalent to conditioning on $L_{I-1}$. The variable $Z$ need not be a
valid one-symbol extension of $L_{I-1}$.

\begin{remark}[Equivalent formulation]
\label{rem:insertional-common-distribution}
Equivalently, every insertional decomposition $(L_0,\ldots,L_n)$
admits random variables $Z_i\in(\{0,1\}^\ell)^i$, jointly distributed with the
decomposition and independently of $I$, satisfying
\[
\sum_{i=1}^n\bigl(H(Z_i\mid L_{i-1})-H(L_i\mid L_{i-1})\bigr)
\geq k-H(L)
\]
and making $(I,Z_I,L_{I-1})$ and $(I,L_I,L_{I-1})$
$(T,\varepsilon)$ nonuniformly indistinguishable.
The family is chosen once for all distinguishing circuits.

Given $Z$, draw each $Z_i$ conditionally on the decomposition using the
distribution of $Z$ given $I=i$ and $L_{i-1}$, with fresh randomness
independent of $I$. On zero-probability conditioning events, choose any
distribution on $(\{0,1\}^\ell)^i$. Conversely, given such a family, take $Z=Z_I$
with $I$ independent of the decomposition and the family.
These constructions preserve the joint distributions
being compared and give
\begin{align*}
H(Z\mid I,L_{I-1})&=\frac1n\sum_{i=1}^nH(Z_i\mid L_{i-1}),\\
H(L_I\mid I,L_{I-1})&=\frac1n\sum_{i=1}^nH(L_i\mid L_{i-1}).
\end{align*}
Thus both formulations have the same entropy gap, circuit size, and
distinguishing error.
\end{remark}

\subsection{Characterization of Insertional Samplability}
\label{subsec:insertional-equivalence}

\begin{theorem}[Characterization of stateless insertional sampling]
\label{thm:insertional-characterization}
There exist a universal constant $c_0\in(0,1]$ and a polynomial $p$ with
nonnegative coefficients and constant term at least one such that the
following holds. Let $n,\ell\geq1$ and let $L$ take values in
$(\{0,1\}^\ell)^n$. Let $t\geq1$,
$0<\epsilon,\eta,\sigma<1$, and $\delta\geq\epsilon$.
\begin{enumerate}
    \item If $L$ has $(t,\epsilon)$ insertional pseudoentropy at least
    $H(L)+\delta$, then, for every $B\geq1$ satisfying
    \[
    B\leq\frac{\min\{t^{c_0},\,\sigma/(n\epsilon)\}}
                   {p(n,2^\ell,1/\eta,\log(n/\sigma))},
    \]
    every size-$B$ stateless insertional sampler $C$ over $\{0,1\}^\ell$ satisfies
    \[
    \kldiv{L}{C(1^n)}>\delta-\sigma-n\eta.
    \]
    \item If the $(t,\epsilon)$ insertional
    pseudoentropy of $L$ is at most $H(L)+\delta-\epsilon$, then there
    exists a size-$B$ stateless insertional sampler $C$ over $\{0,1\}^\ell$ with
    \[
    B\leq p(t,n,2^\ell,1/\epsilon),
    \qquad \kldiv{L}{C(1^n)}\leq\delta.
    \]
\end{enumerate}
\end{theorem}

\begin{remark}
If $\delta<\epsilon$ and some $B\geq1$ satisfies the size bound in item~1, then $p\geq1$ forces $\sigma\geq n\epsilon>\delta$, so the claimed lower bound is negative and the inequality follows from nonnegativity of KL-divergence. If no such $B$ exists, item~1 is vacuous.
\end{remark}

\begin{lemma}[Distinct one-symbol insertions]
\label{lem:subexp-insertion-actions}
Let $i,\ell\geq1$ and $w\in(\{0,1\}^\ell)^{i-1}$. The set
\[
\mathcal E_i(w)=\{z\in(\{0,1\}^\ell)^i:
w\text{ is a subsequence of }z\}
\]
has exactly $K_i=i(2^\ell-1)+1$ elements and admits a bijection
$\mathsf{Ext}_i(w,\cdot):[K_i]\to\mathcal E_i(w)$.
Given $(i,\ell,w,k)$, one can compute $\mathsf{Ext}_i(w,k)$ in
$\operatorname{poly}(i,2^\ell)$ time. Given $(i,\ell,w,z)$ with
$z\in\mathcal E_i(w)$, one can compute the unique $k\in[K_i]$ satisfying
$\mathsf{Ext}_i(w,k)=z$ in the same time bound.
\end{lemma}
\begin{proof}
We count each resulting string once by choosing the rightmost position
at which its inserted symbol can be placed. For $a\in\{0,1\}^\ell$ and
$j\in[i]$, write
\[
\mathsf{Ins}(w,a,j)=(w_1,\ldots,w_{j-1},a,w_j,\ldots,w_{i-1}).
\]
For $a,b\in\{0,1\}^\ell$ and $1\leq j<k\leq i$, comparison of coordinates
$j,\ldots,k$ gives
\[
\mathsf{Ins}(w,a,j)=\mathsf{Ins}(w,b,k)
\quad\Longleftrightarrow\quad
a=w_j=\cdots=w_{k-1}=b.
\]
Retain all pairs $(a,j)$ with $j=i$, and those with $j<i$ and $a\ne w_j$.
By the displayed equivalence, these are exactly the rightmost
representations, one for each output. Hence
\[
|\mathcal E_i(w)|=2^\ell+(i-1)(2^\ell-1)=i(2^\ell-1)+1.
\]
Order the retained pairs by increasing $j$, then lexicographically by $a$.
Define $\mathsf{Ext}_i(w,k)$ by inserting the symbol at the position
specified by the $k$-th pair. To invert this map given $w$ and $z$,
enumerate the retained pairs and return the index of the unique pair
whose insertion produces $z$. Both algorithms examine at most
$i2^\ell$ pairs and use $O(i\ell)$ time per pair, giving the claimed
$\operatorname{poly}(i,2^\ell)$ bound.
\end{proof}

\begin{lemma}[Conditional and final KL-divergence]
\label{lem:insertional-path-kl}
Let $n,\ell\geq1$ and let $L$ be distributed over $(\{0,1\}^\ell)^n$.
For every sequence $C=(C_1,\ldots,C_n)$ of valid insertion circuits,
\[
\inf_{(L_0,\ldots,L_n)}
\sum_{i=1}^n\kldiv{(L_{i-1},L_i)}{(L_{i-1},C_i(L_{i-1}))}
=\kldiv{L}{C(1^n)},
\]
where the infimum is over insertional decompositions of $L$.
Moreover, if $\kldiv{L}{C(1^n)}<\infty$, there exists an insertional
decomposition $(L_0,\ldots,L_n)$ of $L$ that forms a Markov chain and
satisfies
\[
\sum_{i=1}^n\kldiv{(L_{i-1},L_i)}{(L_{i-1},C_i(L_{i-1}))}
=\kldiv{L}{C(1^n)}.
\]
\end{lemma}
\begin{proof}
Let $(L_0,\ldots,L_n)$ be any insertional decomposition of $L$,
and $X_0=\emptyString$, $X_i\gets C_i(X_{i-1})$ with fresh independent
coins, so $X_n\equiv C(1^n)$.
Then,
\begin{align*}
\kldiv{L}{C(1^n)}
&=\kldiv{L_n}{X_n}\\
&\leq\kldiv{(L_{n-1},L_n)}{(X_{n-1},X_n)}\\
&=\kldiv{L_{n-1}}{X_{n-1}}
+\E_{x\sim L_{n-1}}
\kldiv{L_n\mid_{L_{n-1}=x}}{C_n(x)}\\
&\leq
\sum_{i=1}^n\E_{x\sim L_{i-1}}
\kldiv{L_i\mid_{L_{i-1}=x}}{C_i(x)}\\
&=\sum_{i=1}^n
\kldiv{(L_{i-1},L_i)}{(L_{i-1},C_i(L_{i-1}))}.
\end{align*}
The first inequality is \Cref{fact:DPI} applied to the map that keeps
the second coordinate. The next equality is \Cref{fact:kl-chain-rule}:
given $X_{n-1}=x$, the distribution of $X_n$ is $C_n(x)$.
Repeating these two steps down to $i=1$ gives the sum, since
$\kldiv{L_0}{X_0}=0$.
The last equality applies the same chain rule to two pairs with the
same first-coordinate distribution $L_{i-1}$.
Since the decomposition was arbitrary, the final KL divergence is at
most the infimum in the statement.

Write $Q(x_0,\ldots,x_n)=\Pr[X_0=x_0,\ldots,X_n=x_n]$ and
$Q_n(z)=\Pr[X_n=z]$.
For the reverse inequality, assume $\kldiv{L}{Q_n}<\infty$ and define
\[
M(x_0,\ldots,x_n)=
\begin{cases}
Q(x_0,\ldots,x_n)\dfrac{\Pr[L=x_n]}{Q_n(x_n)},&Q_n(x_n)>0,\\
0,&Q_n(x_n)=0.
\end{cases}
\]
For $Q_n(z)>0$,
\begin{align*}
\sum_{x_0,\ldots,x_{n-1}}M(x_0,\ldots,x_{n-1},z)
&=\frac{\Pr[L=z]}{Q_n(z)}
\sum_{x_0,\ldots,x_{n-1}}Q(x_0,\ldots,x_{n-1},z)\\
&=\Pr[L=z].
\end{align*}
Both sides are zero otherwise, since $\supp(L)\subseteq\supp(Q_n)$.
Thus $(M_0,\ldots,M_n)\sim M$ satisfies $M_n\equiv L$.
Since $\supp(M)\subseteq\supp(Q)$, $M_0=\emptyString$ and every step
inserts one symbol.

To compute the conditional probabilities, for $i\in\{0,\ldots,n\}$ set
\[
h_i(w)=\sum_{z:\,Q_n(z)>0}
\Pr[X_n=z\mid X_i=w]\frac{\Pr[L=z]}{Q_n(z)}
\qquad(w\in\supp(X_i)),
\]
and set $h_i(w)=0$ otherwise. The Markov property of $(X_0,\ldots,X_n)$
gives
\begin{align*}
&\Pr[M_0=x_0,\ldots,M_i=x_i]\\
&\quad=\sum_{z:\,Q_n(z)>0}
\Pr[X_0=x_0,\ldots,X_i=x_i,X_n=z]
\frac{\Pr[L=z]}{Q_n(z)}\\
&\quad=\Pr[X_0=x_0,\ldots,X_i=x_i]h_i(x_i).
\end{align*}
For $i\in[n]$ and every tuple with
$\Pr[M_0=x_0,\ldots,M_{i-1}=x_{i-1}]>0$, division gives
\begin{align*}
&\Pr[M_i=z\mid M_0=x_0,\ldots,M_{i-1}=x_{i-1}]\\
&\quad=\Pr[C_i(x_{i-1})=z]\frac{h_i(z)}{h_{i-1}(x_{i-1})}\\
&\quad=\Pr[M_i=z\mid M_{i-1}=x_{i-1}].
\end{align*}
The ratio depends on the preceding tuple only through $x_{i-1}$, so the
last equality holds and $(M_0,\ldots,M_n)$ is a Markov insertional
decomposition of $L$.
By \Cref{fact:kl-chain-rule-markov},
\begin{align*}
&\sum_{i=1}^n
\kldiv{(M_{i-1},M_i)}{(M_{i-1},C_i(M_{i-1}))}\\
&\quad=\kldiv{M}{Q}\\
&\quad=\sum_{(x_0,\ldots,x_n)\in\supp(M)}
M(x_0,\ldots,x_n)\log\frac{\Pr[L=x_n]}{Q_n(x_n)}\\
&\quad=\sum_{z\in\supp(L)}\Pr[L=z]\log\frac{\Pr[L=z]}{Q_n(z)}\\
&\quad=\kldiv{L}{C(1^n)}.
\end{align*}
This decomposition attains the infimum.
\end{proof}

\begin{lemma}[Insertional pseudoentropy implies conditional sampling hardness]
\label{lemma:insertional-PE-implies-conditional-hardness}
Under the hypotheses and size condition of the first part of
\Cref{thm:insertional-characterization}, every insertional decomposition
$(L_0,\ldots,L_n)$ of $L$ and every size-$B$ stateless insertional
sampler $C$ over $\{0,1\}^\ell$ satisfy
\[
\E_{i\gets[n]}
\kldiv{(L_{i-1},L_i)}{(L_{i-1},C_i(L_{i-1}))}
>\frac{\delta-\sigma}{n}-\eta.
\]
\end{lemma}

\begin{proof}
Fix an insertional decomposition $(L_0,\ldots,L_n)$ and a sampler $C$, and put
\[
d_i=\kldiv{(L_{i-1},L_i)}{(L_{i-1},C_i(L_{i-1}))},
\qquad d=\sum_{i=1}^n d_i.
\]
If $d=\infty$, the claim follows, so assume $d<\infty$.
By \Cref{def:insertionalPseudoentropy} and \Cref{rem:insertional-common-distribution},
choose a family $Z_i\in(\{0,1\}^\ell)^i$ such that
\[
\sum_{i=1}^n H(Z_i\mid L_{i-1})
\geq\sum_{i=1}^n H(L_i\mid L_{i-1})+\delta
\]
and $(I,Z_I,L_{I-1})$ and $(I,L_I,L_{I-1})$ are
$(t,\epsilon)$ nonuniformly indistinguishable, where $I$ is uniform on
$[n]$ and independent of the family and decomposition.
We estimate the sampler's output probabilities and test their negative
logarithms. The expected negative logarithm will be at least $H(Z_i\mid L_{i-1})$
on $Z_i$ and at most $H(L_i\mid L_{i-1})+d_i+\eta/4$ on $L_i$.
Indistinguishability compares their averages over the uniform index $I$.

Let $\mathcal E_i(w)$ and $K_i=i(2^\ell-1)+1$ be as in
\Cref{lem:subexp-insertion-actions}, and set
\[
q=n2^\ell,\qquad M=\lceil32q/\eta\rceil,\qquad
\theta=2^{-\lceil\log(32/\eta)\rceil},\qquad
\Lambda=n\ell+\log(1/\theta).
\]
On input $(i,z,w)$, draw $M$ samples from $C_i(w)$ with independent
coins, also independent of the decomposition and auxiliary family.
Let $N_z$ count the samples equal to $z$. Define
\[
\widehat P_w(z)=
\begin{cases}
\dfrac{N_z+1}{M+K_i},&z\in\mathcal E_i(w),\\
0,&z\notin\mathcal E_i(w),
\end{cases}
\qquad
R_w(z)=(1-\theta)\widehat P_w(z)+\frac{\theta}{2^{i\ell}}.
\]
Since $\sum_{z\in\mathcal E_i(w)}N_z=M$, both $\widehat P_w$ and $R_w$
are probability distributions. The uniform term is needed because $Z_i$
need not belong to $\mathcal E_i(L_{i-1})$. For every length-$i$ string $z$,
\[
\frac{\theta}{2^{i\ell}}\leq R_w(z)\leq1,
\qquad 0\leq-\log R_w(z)\leq\Lambda.
\]

Set $b=\lceil\log(16\Lambda/\eta)\rceil$ and $\rho=2^{-b}$.
The circuit $D$ accepts with probability
\[
\rho\left\lfloor\frac{-\log R_w(z)}{\Lambda\rho}\right\rfloor.
\]
For each $i$, store the displayed floor for the $M+1$ valid-extension
counts $N_z=0,\ldots,M$ and for the single invalid-extension case.
This gives $n(M+2)$ entries, each an integer in $\{0,\ldots,2^b\}$
stored using $b+1$ bits. Draw a fresh uniform $b$-bit integer and accept
exactly when it is smaller than the stored integer.
The circuit contains $M$ copies of each $C_j$; it selects the count from
the branch $j=i$, whose copies receive $w$; the other branches receive
fixed strings of the required lengths. Membership in
$\mathcal E_i(w)$ is tested by deleting each position of $z$ and comparing
the resulting string with $w$. Selection, counting, these comparisons,
and table lookup require $\operatorname{poly}(n,M,\ell,b)$ additional gates.
Consequently, for a universal polynomial $A$,
\[
|D|\leq nMB+\operatorname{poly}(n,M,\ell,b)
\leq B A(n,2^\ell,1/\eta),
\]
where the last inequality uses $B\geq1$ and the choices of $M,b$.
Choose the polynomial $p$ in \Cref{thm:insertional-characterization}
large enough that
$p(n,2^\ell,1/\eta,\log(n/\sigma))\geq\max\{A(n,2^\ell,1/\eta),2\Lambda\}$.
The assumed size bound and $c_0\leq1$ give
\[
|D|\leq BA(n,2^\ell,1/\eta)\leq t^{c_0}\leq t,
\qquad n\Lambda\epsilon\leq\sigma/2,
\qquad \Lambda\rho\leq\eta/16.
\]

We first bound the expected negative logarithm on $L_i$.
For $w\in\supp(L_{i-1})$, apply
\Cref{claim:bound-expected-kl-laplace} on $\mathcal E_i(w)$ with
$P=C_i(w)$, $Q=L_i\mid_{L_{i-1}=w}$, and $M$ samples. Since
$R_w(z)\geq(1-\theta)\widehat P_w(z)$,
\begin{align*}
\E\kldiv{L_i\mid_{L_{i-1}=w}}{R_w}
&\leq\E\kldiv{L_i\mid_{L_{i-1}=w}}{\widehat P_w}
-\log(1-\theta)\\
&\leq\kldiv{L_i\mid_{L_{i-1}=w}}{C_i(w)}
+\log\frac{M+K_i}{M+1}-\log(1-\theta).
\end{align*}
Here expectations are over the $M$ samples. Since $K_i\leq q$ and
$\theta\leq\eta/32$, the parameter choices give
\[
\log\frac{M+K_i}{M+1}-\log(1-\theta)
<\eta/4.
\]
Drawing the samples independently of $L_i$ conditional on $L_{i-1}$ and
applying \Cref{fact:kl-chain-rule} to $d_i$ gives
\begin{align*}
\E[-\log R_{L_{i-1}}(L_i)]
&=H(L_i\mid L_{i-1})
+\E_{w\sim L_{i-1}}\E\kldiv{L_i\mid_{L_{i-1}=w}}{R_w}\\
&\leq H(L_i\mid L_{i-1})+d_i+\eta/4.
\end{align*}

The same samples are independent of $Z_i$ conditional on $L_{i-1}$.
For every $w\in\supp(L_{i-1})$ and every value of $R_w$,
\begin{align*}
\E_{z\sim Z_i\mid_{L_{i-1}=w}}[-\log R_w(z)]
&=H(Z_i\mid L_{i-1}=w)
+\kldiv{Z_i\mid_{L_{i-1}=w}}{R_w}\\
&\geq H(Z_i\mid L_{i-1}=w).
\end{align*}
For $V_i=L_i$ or $V_i=Z_i$, the rounding gives
\[
\left|\Lambda\Pr[D(i,V_i,L_{i-1})=1]
-\E[-\log R_{L_{i-1}}(V_i)]\right|\leq\Lambda\rho.
\]
Here the expectations include the fresh samples drawn by $D$.
Since $|D|\leq t$, indistinguishability gives
\[
\left|\sum_i\Pr[D(i,Z_i,L_{i-1})=1]
-\sum_i\Pr[D(i,L_i,L_{i-1})=1]\right|\leq n\epsilon.
\]
Combining these inequalities,
\begin{align*}
\sum_i H(L_i\mid L_{i-1})+\delta
&\leq\sum_i H(Z_i\mid L_{i-1})\\
&\leq\sum_i\E[-\log R_{L_{i-1}}(Z_i)]\\
&\leq\Lambda\sum_i\Pr[D(i,Z_i,L_{i-1})=1]+n\Lambda\rho\\
&\leq\Lambda\sum_i\Pr[D(i,L_i,L_{i-1})=1]
+n\Lambda\epsilon+n\Lambda\rho\\
&\leq\sum_i\E[-\log R_{L_{i-1}}(L_i)]
+n\Lambda\epsilon+2n\Lambda\rho\\
&\leq\sum_i H(L_i\mid L_{i-1})+d+n\eta/4
+n\Lambda\epsilon+2n\Lambda\rho\\
&\leq\sum_i H(L_i\mid L_{i-1})+d+\sigma/2+3n\eta/8.
\end{align*}
Subtracting $\sum_iH(L_i\mid L_{i-1})$ from both sides gives
$d > \delta-\sigma-n\eta$.
Dividing by $n$ proves the claim.
\end{proof}

\begin{lemma}[Small insertional pseudoentropy gives conditional samplers]
\label{lemma:small-insertional-PE-gives-conditional-samplers}
Under the hypotheses of the second part of
\Cref{thm:insertional-characterization}, there are an insertional
decomposition $(L_0,\ldots,L_n)$ of $L$ and a size-$B$ stateless
insertional sampler $C$ over $\{0,1\}^\ell$, with
$B\leq p(t,n,2^\ell,1/\epsilon)$, such that
\[
\E_{i\gets[n]}
\kldiv{(L_{i-1},L_i)}{(L_{i-1},C_i(L_{i-1}))}
\leq\frac{\delta}{n}.
\]
\end{lemma}

\begin{proof}
By assumption, the $(t,\epsilon)$ insertional pseudoentropy of $L$ is
at most $H(L)+\delta-\epsilon$.
Indexing the valid extensions in \Cref{lem:subexp-insertion-actions}
gives alphabets of size at most $n2^\ell$ for applying item~2 of
\Cref{thm:formal-characterizing-pseudoentropy-nonuniform}.
We bound the sum of the indices' conditional pseudoentropy gaps by
combining their maximizing auxiliary variables into one variable in
\Cref{def:insertionalPseudoentropy}.
Set
\[
k_*=H(L)+\delta-\epsilon/2,
\qquad \eta'=\epsilon/(4n),
\qquad 2n\eta'=\epsilon/2.
\]
Since $H(L)+\delta-\epsilon<k_*$, \Cref{def:insertionalPseudoentropy}
gives an insertional decomposition $(L_0,\ldots,L_n)$ for which every
$Z$ satisfying the length and $(t,\epsilon)$-indistinguishability
conditions in that definition satisfies
\[
nH(Z\mid I,L_{I-1})-\sum_{i=1}^nH(L_i\mid L_{i-1})
<k_*-H(L)=\delta-\epsilon/2.
\]
Here $I$ is uniform on $[n]$ and independent of the decomposition.

For each $i$, let $K_i=i(2^\ell-1)+1$ and let
$\mathsf{Ext}_i(w,\cdot):[K_i]\to\mathcal E_i(w)$ be the bijection in
\Cref{lem:subexp-insertion-actions}.
Define $A_i\in[K_i]$ by $L_i=\mathsf{Ext}_i(L_{i-1},A_i)$.
Since the map is a bijection for each fixed $L_{i-1}=w$,
\[
H(A_i\mid L_{i-1})=H(L_i\mid L_{i-1}).
\]
Let $R(n,2^\ell)$ be a fixed polynomial bounding the circuit size of
each $\mathsf{Ext}_i$ and the overhead for composing it with a circuit
whose index input is fixed to $i$, and put $t'=t+R(n,2^\ell)$.
Such a polynomial exists by the algorithms in
\Cref{lem:subexp-insertion-actions}.

Let $h_i$ be the maximum $(t',\eta')$ conditional pseudoentropy of
$A_i$ given $L_{i-1}$ over the alphabet $[K_i]$, and put
\[
g_i=h_i-H(A_i\mid L_{i-1})\geq0.
\]
Nonnegativity follows by using $A_i$ itself as an auxiliary variable.
We will show $\sum_i g_i<\delta-\epsilon/2$.
By the compactness observation after
\Cref{def:conditional-pseudoentropy-nonuniform}, choose maximizing
$A_i'\in[K_i]$ such that $H(A_i'\mid L_{i-1})=h_i$ and
$(L_{i-1},A_i')$ and $(L_{i-1},A_i)$ are
$(t',\eta')$ nonuniformly indistinguishable.
Realize these pairs jointly with $(L_0,\ldots,L_n)$ by setting
\begin{align*}
&\Pr[A_1'=a_1,\ldots,A_n'=a_n
\mid L_0=w_0,\ldots,L_n=w_n]\\
&\qquad=\prod_{i=1}^n
\Pr[A_i'=a_i\mid L_{i-1}=w_{i-1}]
\end{align*}
for every tuple $(w_0,\ldots,w_n)$ of positive probability, using the
chosen pair distributions on the right. This preserves each distribution
$(L_{i-1},A_i')$. Put
\[
Z_i=\mathsf{Ext}_i(L_{i-1},A_i'),
\qquad H(Z_i\mid L_{i-1})=H(A_i'\mid L_{i-1})=h_i,
\]
choose $I$ independently of this family and the decomposition, and set
$Z=Z_I$.

Fix any Boolean circuit $D$ of size at most $t$. For each fixed $i$,
the circuit with inputs $w\in(\{0,1\}^\ell)^{i-1}$ and $a\in[K_i]$
given by
\[
D_i(w,a)=D(i,\mathsf{Ext}_i(w,a),w)
\]
has size at most $t+R(n,2^\ell)=t'$.
The defining indistinguishability of $A_i'$ gives
\begin{align*}
&\left|\Pr[D(I,Z,L_{I-1})=1]
-\Pr[D(I,L_I,L_{I-1})=1]\right|\\
&\quad=\left|\frac1n\sum_{i=1}^n
\left(\Pr[D_i(L_{i-1},A_i')=1]
-\Pr[D_i(L_{i-1},A_i)=1]\right)\right|\\
&\quad\leq\frac1n\sum_{i=1}^n
\left|\Pr[D_i(L_{i-1},A_i')=1]
-\Pr[D_i(L_{i-1},A_i)=1]\right|\\
&\quad\leq\eta'\leq\epsilon.
\end{align*}
The family $(Z_1,\ldots,Z_n)$ was fixed before $D$, and
$Z\in(\{0,1\}^\ell)^i$ when $I=i$. Thus $Z$ satisfies the length and
indistinguishability conditions for the chosen decomposition, so
\begin{align*}
\sum_{i=1}^n g_i
&=\sum_{i=1}^n h_i-\sum_{i=1}^n H(A_i\mid L_{i-1})\\
&=\sum_{i=1}^n H(Z_i\mid L_{i-1})
-\sum_{i=1}^n H(L_i\mid L_{i-1})\\
&=nH(Z\mid I,L_{I-1})-\sum_{i=1}^n H(L_i\mid L_{i-1})\\
&<k_*-H(L)=\delta-\epsilon/2.
\end{align*}

Apply item~2 of
\Cref{thm:formal-characterizing-pseudoentropy-nonuniform} for each $i$
with $X=L_{i-1}$, $B=A_i$, $m=m_i=(i-1)\ell\geq0$, $q=K_i\geq2$,
circuit-size parameter $t'$, gap bound $g_i$, and $\eta=\eta'$.
It gives a randomized circuit $S_i:\{0,1\}^{m_i}\to[K_i]$ satisfying
\begin{align*}
|S_i|&\leq p_0(t',m_i,K_i,1/\eta'),\\
\kldiv{(L_{i-1},A_i)}{(L_{i-1},S_i(L_{i-1}))}
&\leq g_i+2\eta'.
\end{align*}
Set $C_i(w)=\mathsf{Ext}_i(w,S_i(w))$, with fresh independent coins
at each step. Since $S_i(w)\in[K_i]$, $C_i(w)\in\mathcal E_i(w)$
for every input $w\in(\{0,1\}^\ell)^{i-1}$.
The bijection $(w,a)\mapsto(w,\mathsf{Ext}_i(w,a))$ gives
\begin{align*}
\kldiv{(L_{i-1},L_i)}{(L_{i-1},C_i(L_{i-1}))}
&=\kldiv{(L_{i-1},A_i)}{(L_{i-1},S_i(L_{i-1}))}\\
&\leq g_i+2\eta'.
\end{align*}
Hence
\begin{align*}
\sum_{i=1}^n\kldiv{(L_{i-1},L_i)}{(L_{i-1},C_i(L_{i-1}))}
&\leq\sum_{i=1}^n g_i+2n\eta'\\
&<\delta-\epsilon/2+\epsilon/2=\delta.
\end{align*}
Dividing by $n$ gives the stated average bound.

Finally, $m_i\leq n2^\ell$, $K_i\leq n2^\ell$, and
$1/\eta'=4n/\epsilon$. Since $p_0$ has nonnegative coefficients,
\begin{align*}
|C_i|&\leq |S_i|+R(n,2^\ell)\\
&\leq p_0(t+R(n,2^\ell),n2^\ell,n2^\ell,4n/\epsilon)+R(n,2^\ell).
\end{align*}
Choose the polynomial $p$ in \Cref{thm:insertional-characterization}
large enough to dominate this quantity at $(t,n,2^\ell,1/\epsilon)$
and the bound required in its first part. Then
$B:=\max_i|C_i|\leq p(t,n,2^\ell,1/\epsilon)$.
\end{proof}

\begin{proof}[Proof of \Cref{thm:insertional-characterization}]
For the first part, fix a sampler $C$ satisfying the size bound.

Write $d=\kldiv{L}{C(1^n)}$. If $d=\infty$, the claim follows. Otherwise,
\Cref{lem:insertional-path-kl} gives an insertional decomposition
$(L_0,\ldots,L_n)$ of $L$ whose sum of conditional KL-divergences
from $C$ equals $d$.
Applying \Cref{lemma:insertional-PE-implies-conditional-hardness} to
this decomposition gives
\begin{align*}
\kldiv{L}{C(1^n)}
&=\sum_{i=1}^n\kldiv{(L_{i-1},L_i)}{(L_{i-1},C_i(L_{i-1}))}\\
&>\delta-\sigma-n\eta.
\end{align*}

For the second part, take the decomposition and sampler from
\Cref{lemma:small-insertional-PE-gives-conditional-samplers}.
The sampler has the required size, and \Cref{lem:insertional-path-kl}
gives
\[
\kldiv{L}{C(1^n)}
\leq\sum_i\kldiv{(L_{i-1},L_i)}{(L_{i-1},C_i(L_{i-1}))}
\leq\delta.
\]
\end{proof}

Finally, insertion does not recover all efficient sampling:
\Cref{thm:sm-insertional-separation} proves a separation under
$2^{\gamma\lambda^\alpha}$-secure one-way functions, for constants
$\gamma>0$ and $1/2<\alpha<1$. This construction uses $2t$ tokens over
an alphabet of size $2t\cdot 2^{2t}$.

\section{Insertionally-Samplable PRGs}
A \emph{pseudorandom generator} (PRG) is a polynomial-time family
$G:\zo^k\to\zo^m$, with $m>k$, whose output on a uniform seed is
computationally indistinguishable from $U_m$. Here security is against
nonuniform polynomial-size circuits: every such circuit has negligible
distinguishing advantage. The following construction shows that an
insertional sampler can generate pseudorandom bits without retaining
the seed privately.

\begin{theorem}[Insertionally samplable pseudorandom generators]
\label{prop:insertionally-samplable-prg}
Let $G:\{0,1\}^t\to\{0,1\}^{2t^2}$ be a polynomial-time
pseudorandom generator secure against nonuniform polynomial-size
circuits. There exists a polynomial-time pseudorandom generator
$H:\{0,1\}^{3t^2}\to\{0,1\}^{4t^2}$ whose output distribution
admits an exact polynomial-size stateless insertional bit sampler.
\end{theorem}
\begin{proof}
The sampler first generates bits from which it can recover a seed.
After using this seed, it inserts additional random bits so that the
final string conceals which bits were originally generated.

Define $H$ by the following process, with its seed consisting of
the $3t^2$ independent uniform bits used below.
\begin{enumerate}
    \item Append $t^2$ independent uniform bits, obtaining
    $X=X_1\cdots X_{t^2}$. Partition $X$ into $t$ consecutive
    groups of $t$ bits. Define $s_j$ as the parity of the bits in group $j$,
    and put $s=s_1\cdots s_t$. 

    \item For $a=1,\ldots,2t^2$, read $X$ from the first $t^2$
    positions, recompute $s$, and append $G(s)_a$.
    (Following this process, the resulting string is $X\Vert G(s)$.)

    \item For $i=1,\ldots,t^2$, sample fresh independent uniform
    bits $b,r$. Insert $b$ immediately before $X_i$ if $r=0$,
    and immediately after it otherwise.
    Denote the resulting ordered pair by $Z_i$, and put
    $Z=Z_1\Vert\cdots\Vert Z_{t^2}$.
    The final output is $Z\Vert G(s)$.
\end{enumerate}

The process uses $t^2$ random bits in item~1 and $2t^2$ in item~3,
and outputs $4t^2$ bits. It is stateless: throughout item~2, $X$
occupies the first $t^2$ positions. After processing $i-1$ bits in
item~3, the current string is
$$
Z_1\Vert\cdots\Vert Z_{i-1}\Vert
X_i\cdots X_{t^2}\Vert G(s),
$$
so $X_i$ occupies position $2i-1$.
The current length determines the phase and index. Every transition
therefore uses only the current string and fresh coins, and has
polynomial-size circuits.

It remains to prove pseudorandomness. The pairs $Z_i$ are independent
and uniform over $\{00,01,10,11\}$, so $Z$ is uniform.
Conditioned on any $Z=z$, the original bits $X_i$ remain independent:
an equal pair determines its original bit, whereas an unequal pair
leaves its original bit uniform. Hence, if a group contains an
unequal pair, its parity is uniform given $Z=z$.

The probability that all $t$ pairs of a given group are equal is
$2^{-t}$. Except on an event of probability at most $t2^{-t}$,
every group therefore has an unequal pair, making all seed bits
independent and uniform given $Z$. Consequently,
$$
\statDist\bigl((Z,s),(U_{2t^2},U_t)\bigr)\leq t2^{-t},
$$
where the uniforms on the right are independent.
Applying $(z,s)\mapsto z\Vert G(s)$ shows that $H(U_{3t^2})$
is statistically close to $U_{2t^2}\Vert G(U_t)$.
By security of $G$, the latter is computationally indistinguishable
from $U_{4t^2}$, proving security of $H$.
\end{proof}

\section{Separations in Sampling and Pseudoentropy}
\label{sec:separations}

We separate insertional from sequential sampling over a binary alphabet
in \Cref{subsec:insertional-vs-sequential}, and general from insertional
sampling over a larger alphabet in
\Cref{subsec:insertional-vs-ordinary-pseudoentropy}.
In \Cref{subsec:pseudoentropy-separations}, we derive the corresponding
separations between pseudoentropy notions.

\subsection{Separating Insertional Samplers from Sequential Samplers}
\label{subsec:insertional-vs-sequential}

In this section formalize the intuition described in \cite{wu2023limitations} who described a distribution that was hard for sequential samplers to derive. Moreover, we show that the same intuition does not apply to insertional samplers, thereby showing that insertional samplers cover a larger class of distributions compared to their sequential counterparts.

\begin{proposition}[Insertional sampling is stronger than sequential sampling]
\label{thm:insertional-vs-sequential}
Let $f\colon\{0,1\}^{n}\to\{0,1\}^{n}$ be a one-way function
secure against polynomial-size circuits. 
There exists a distribution $L$ over $\zo^{2n}$ such that:
\begin{enumerate}
    \item Every stateless sequential sampler of size $\poly(n)$ satisfies $\kldiv{L}{C(1^{2n})}>\log(n)$, and
    \item There is a $\poly(n)$-size insertional sampler that samples $L$.
\end{enumerate}
\end{proposition}
\begin{proof}
We set $L$ to be the distribution $L=(f(X),X)$ where $X$ is uniform over $\zo^{n}$. 

We now prove the first item. Suppose towards contradiction that for some constant $c>0$ and infinitely many
$n$, a size-$n^c$ stateless sequential sampler $C$
satisfies $\kldiv{L}{C(1^{2n})}\leq\log n$. We show that this invalidates security of the one-way function.
Given $y=f(x)$ the inverter $I$ runs $C$ sequentially on prefix $y$ and outputs the $n$-bit suffix generated by $C$.

For $X\gets U_n$, put $Y=f(X)$ and write $(V,W)\gets C(1^{2n})$,
where $V,W\in\zo^n$. Since $C$ is stateless, $W\mid_{V=y}$ is
distributed as $I(y)$ for every $y\in\supp(V)$. The assumed finite
KL-divergence ensures that $\supp(Y)\subseteq\supp(V)$.
Applying the KL chain rule (\Cref{fact:kl-chain-rule}) to both pairs gives
\begin{align*}
\kldiv{(Y,X)}{(Y,I(Y))}
&=\E_{y\gets Y}\left[\kldiv{X\mid_{Y=y}}{I(y)}\right]\\
&=\kldiv{L}{C(1^{2n})}-\kldiv{Y}{V}
\leq\log n,
\end{align*}
where the inequality uses nonnegativity of KL-divergence.

Let $p=\Pr[f(I(Y))=Y]$ be the inverter's success probability.
The test $T(y,x')=\mathds{1}[f(x')=y]$ maps $(Y,X)$ to the constant
$1$ and $(Y,I(Y))$ to a Bernoulli random variable with parameter $p$.
Thus data processing (\Cref{fact:DPI}) gives
\begin{align*}
\log(1/p)
&=\kldiv{\operatorname{Ber}(1)}{\operatorname{Ber}(p)}\\
&\leq\kldiv{(Y,X)}{(Y,I(Y))}\leq\log n.
\end{align*}
Hence the polynomial-size inverter $I$ succeeds with probability
$p\geq1/n$ for infinitely many $n$, contradicting one-wayness.

We now show the second item by describing an insertional sampler for $L$:
\begin{enumerate}[noitemsep]
    \item For the first $n$ steps, append independent uniform bits,
    obtaining $x\gets U_{n}$.
    \item At step $n+j$, for $j\in[n]$, read $x$ from the input, compute $f(x)$ and output $(j, f(x)_{j})$ (i.e., place the $j$-th bit of $f(x)$ at index $j$.
\end{enumerate}
That this insertional sampler samples $L$ exactly and is efficient is immediate by construction.

\end{proof}

\subsection{Separating General Samplers from Insertional Samplers}
\label{subsec:insertional-vs-ordinary-pseudoentropy}

We construct an efficiently samplable distribution that is hard for
stateless insertional samplers.

\begin{theorem}[General sampling is stronger than insertional sampling]
\label{thm:sm-insertional-separation}
Let $(f_\lambda)_{\lambda\geq1}$ be a length-preserving one-way
function family with $2^{\gamma\lambda^\alpha}$ security against
circuits, for constants $\gamma>0$ and $1/2<\alpha<1$.
For $t\geq2$, let $\ell=2t+\lceil\log(2t)\rceil$.
There exists a distribution $L=L_t$ over $(\zo^\ell)^{2t}$ such that:
\begin{enumerate}
	 \item For every constant $c>0$ and all sufficiently large $t$,
	every stateless insertional sampler $C$ of size at most $2^{ct}$
	satisfies $\kldiv{L}{C(1^{2t})}\geq\gamma t^{2\alpha}-2t$.
    \item There is a $\poly(t)$-size randomized circuit that samples $L$ exactly.
\end{enumerate}
\end{theorem}
\begin{proof}
We first define $L$. Let $q=2^t$, fix distinct
$\beta_1,\ldots,\beta_{2t}\in\mathbb F_q$, and identify
$\mathbb F_q^t$ with $\zo^{t^2}$. For $x,r,y\in\mathbb F_q^t$, put
$P_{x,r}(z)=\sum_{j<t}r_jz^j+\sum_{j<t}x_jz^{t+j}$ and
$Q_y(z)=\sum_{j<t}y_jz^j$.
The distribution $L$ is sampled as follows:
\begin{enumerate}[noitemsep]
    \item Sample independent uniform $X,R\gets\mathbb F_q^t$ and
    compute $Y=f_{t^2}(X)$.
    \item Output $((i,P_{X,R}(\beta_i),Q_Y(\beta_i)))_{i=1}^{2t}$,
    treating each triple as one $\ell$-bit token.
\end{enumerate}
The evaluations of $P_{X,R}$ form a ramp secret-sharing of $X$:
any fixed $t$ shares are uniform and independent of $X$, since the
Vandermonde matrix acting on $R$ is invertible, while all $2t$ shares
recover $X,R$ by interpolation. Intermediate numbers of shares may
reveal partial information about $X$. In contrast, any $t$ evaluations
of $Q_Y$ determine $Y$. Thus we can sample any fixed $t$ tokens given
only the image, but completing them to a valid output reveals a preimage.

Observe that the description of the distribution defines a circuit for sampling from it. This sampler runs in $\poly(t)$ time, proving the second item. Its alphabet is $[2t]\times\mathbb F_q^2$, which has size $2t\,2^{2t}$.

We now prove the first item. Suppose towards contradiction that for
some constant $c>0$ and infinitely many $t$, a stateless insertional
sampler $C$ of size at most $2^{ct}$ has KL error
$d:=\kldiv{L}{C(1^{2t})}$ smaller than $\gamma t^{2\alpha}-2t$.
We show that this invalidates security of the one-way function.

Let $\mathcal S=\supp(L)$ and observe that membership in $\mathcal S$ can be efficiently tested by interpolating to derive $X$ and then checking whether the third entries in the output are consistent with $f(X)$.
We now describe the inverter. Let $b=\lceil\log\binom{2t}{t}\rceil\leq2t$.
Given $y=f_{t^2}(x)$, the inverter $I$ proceeds as follows:
\begin{enumerate}[noitemsep]
    \item Choose a $t$-element subset $J\subseteq[2t]$ by reducing
    a uniform $b$-bit integer modulo $\binom{2t}{t}$ and taking the
    corresponding subset in lexicographic order.
    \item Sample independent uniform $u_j\gets\mathbb F_q$ for $j\in J$.
    Set $W_t=((j,u_j,Q_y(\beta_j)))_{j\in J}$ in increasing tag order.
    \item For $i=t+1,\ldots,2t$, sample $W_i\gets C_i(W_{i-1})$
    using fresh coins.
    \item If $W_{2t}\in\mathcal S$, output the preimage recovered by interpolating the entries that encode $x$; otherwise output $\bot$.
\end{enumerate}
On acceptance, the preserved $t$ evaluations of $Q_y$ agree with
those of $Q_{f_{t^2}(x')}$ for the recovered $x'$. Both polynomials
have degree less than $t$, so $f_{t^2}(x')=y$.
The inverter has size $2^{O(t)}$: it makes $t$ calls to $C$ and
otherwise uses $\poly(t)$ computation.

It remains to lower-bound its success probability. Below, the strings
$W_t,\ldots,W_{2t}$ are generated on a random challenge
$Y=f_{t^2}(X)$ with $X\gets U_{t^2}$.
By \Cref{lem:insertional-path-kl}, there is a Markov insertional
decomposition $(L_0,\ldots,L_{2t})$ of $L$ such that, writing
$d_i=\kldiv{(L_{i-1},L_i)}{(L_{i-1},C_i(L_{i-1}))}$,
we have $\sum_{i=1}^{2t}d_i=d$.

For a fixed set $J$ of $t$ tags, write $L|_J$ for the restriction
of $L$ to those tags. The secret-sharing property implies that
$W_t$, conditioned on choosing $J$, is distributed as $L|_J$.
Moreover, each $J$ is chosen with probability at least $2^{-b}$.
If $w$ has tag set $J$, the event $L_t=w$ implies $L|_J=w$.
Consequently, for every $w\in\supp(L_t)$,
\[
\Pr[L_t=w]\leq\Pr[L|_J=w]\leq2^b\Pr[W_t=w],
\]
and hence $\kldiv{L_t}{W_t}\leq b$. This comparison allows the
indices in $L_t$ to depend on the final string; it uses secrecy only
for each fixed $J$.

Since $C$ is stateless, the inverter continues from $W_t$ using only
the current string. The KL chain rule (\Cref{fact:kl-chain-rule}) and
nonnegativity of the $d_i$ therefore give
\begin{equation*}
\kldiv{(L_t,\ldots,L_{2t})}{(W_t,\ldots,W_{2t})}
=\kldiv{L_t}{W_t}+\sum_{i=t+1}^{2t}d_i\leq b+d.
\end{equation*}
Let $p=\Pr[W_{2t}\in\mathcal S]$ be the inverter's success probability.
The support test accepts $L_{2t}$ with probability one and $W_{2t}$
with probability $p$. Thus data processing (\Cref{fact:DPI}) gives
\begin{equation*}
\log(1/p)
=\kldiv{\operatorname{Ber}(1)}{\operatorname{Ber}(p)}
\leq b+d<\gamma t^{2\alpha}.
\end{equation*}
Hence $p>2^{-\gamma t^{2\alpha}}$. Since $2\alpha>1$, the inverter's
size $2^{O(t)}$ is at most $2^{\gamma t^{2\alpha}}$ for all sufficiently
large $t$. This contradicts the assumed security at input length
$\lambda=t^2$ for infinitely many $t$, proving the second item.
\end{proof}

\subsection{Separating Pseudoentropy Notions}
\label{subsec:pseudoentropy-separations}

We now combine the sampler separations with
\Cref{thm:sequential-characterization,thm:insertional-characterization}
to separate the corresponding pseudoentropy notions. For each notion,
the pseudoentropy gap is its pseudoentropy minus the Shannon entropy
of the distribution.

\begin{corollary}[Separating insertional from next-block pseudoentropy gaps]
\label{cor:insertional-vs-next-block-pseudoentropy}
Let $(f_n)_{n\geq1}$ be a length-preserving one-way function family
secure against polynomial-size circuits. For $X\gets U_n$, let
$L=(f_n(X),X)$, viewed as $2n$ binary tokens.
For every sufficiently large constant $c$ and all sufficiently large $n$:
\begin{enumerate}
    \item The $(n^c,n^{-c})$ insertional pseudoentropy gap of $L$
    is at most $1/n$.
    \item The $(n^c,n^{-c})$ next-block pseudoentropy gap of $L$
    is at least $\tfrac12\log n$.
\end{enumerate}
\end{corollary}
\begin{proof}
Since $L$ includes the uniform preimage, $H(L)=n$.
For the first item, use the fixed insertion order from the proof of
\Cref{thm:insertional-vs-sequential}, and let $(L_0,\ldots,L_{2n})$
be its intermediate strings. Write $A_i(w)$ for the possible successors
of $w$: for $i\leq n$, these are $w0,w1$; for $i>n$, there is only
the prescribed insertion of an image bit. A circuit of size $T_0(n)$
checks whether $z\in A_i(w)$, for some polynomial $T_0$.
Fix $c\geq4$ large enough that $T_0(n)\leq n^c$ for all sufficiently
large $n$, and set $\varepsilon=n^{-c}$.

Let $I$ be uniform on $[2n]$ and let $Z$ be an auxiliary variable
satisfying the $(n^c,\varepsilon)$ indistinguishability condition in
\Cref{def:insertionalPseudoentropy} for this decomposition.
Put $a=\Pr[Z\notin A_I(L_{I-1})]$.
The test above gives $a\leq\varepsilon$.
Conditioning on whether this event occurs gives
\begin{align*}
H(Z\mid I,L_{I-1})
&\leq h(a)+\E[\log|A_I(L_{I-1})|]+2na\\
&\leq h(\varepsilon)+\tfrac12+2n\varepsilon.
\end{align*}
Since $H(L)=n$, the insertional pseudoentropy gap is at most
$2n h(\varepsilon)+4n^2\varepsilon$.
Since $h(\varepsilon)=O(\varepsilon\log(1/\varepsilon))$, this is
$O(n^{2-c})\leq1/n$ for all sufficiently large $n$, proving the first item.

For the second item, suppose that for some $c>0$ and infinitely many
$n$, the $(n^c,n^{-c})$ next-block pseudoentropy gap is less than the
claimed bound. For all sufficiently large such $n$,
$\tfrac12\log n\leq\log n-n^{-c}$.
Part 2 of \Cref{thm:sequential-characterization}, with
$\delta=\log n$ and $\epsilon=n^{-c}$, gives a stateless sequential
sampler of size $\poly(n^c,2n,2,n^c)=\poly(n)$ and KL error at most
$\log n$, contradicting \Cref{thm:insertional-vs-sequential}.
\end{proof}

\begin{corollary}[Separating insertional from ordinary pseudoentropy gaps]
\label{cor:insertional-vs-ordinary-pseudoentropy}
Let $(f_\lambda)_{\lambda\geq1}$ be a length-preserving one-way
function family with $2^{\gamma\lambda^\alpha}$ security against
circuits, for constants $\gamma>0$ and $1/2<\alpha<1$.
For $t\geq2$, let $\ell=2t+\lceil\log(2t)\rceil$.
There is a distribution $L=L_t$ over $(\zo^\ell)^{2t}$, samplable
in $\poly(t)$ time, such that for every sufficiently large constant $c$
and all sufficiently large $t$:
\begin{enumerate}
    \item The ordinary $(t^c,t^{-c})$ pseudoentropy gap of $L$
    is at most $1/t$.
    \item The $(t^c,t^{-c})$ insertional pseudoentropy gap of $L$
    is at least $\tfrac\gamma2 t^{2\alpha}$.
\end{enumerate}
\end{corollary}
\begin{proof}
Use the distribution constructed in \Cref{thm:sm-insertional-separation}.
Every output determines the independent uniform $X,R$, so $L$ is
uniform on its support $\mathcal S$ of size $2^{2t^2}$ and $H(L)=2t^2$.
Let $T_0(t)$ be a polynomial bound on the size of the support test in
that proof. Fix $c\geq4$ large enough that $T_0(t)\leq t^c$ for all
sufficiently large $t$, and set $\varepsilon=t^{-c}$.
If $V$ is $(t^c,\varepsilon)$-indistinguishable from $L$,
then $a=\Pr[V\notin\mathcal S]\leq\varepsilon$.
Conditioning on membership in $\mathcal S$ gives
\[
H(V)\leq h(a)+(1-a)2t^2+a(2t\ell)
\leq H(L)+h(\varepsilon)+2t\ell\varepsilon,
\]
so the ordinary pseudoentropy gap is at most
$h(\varepsilon)+2t\ell\varepsilon=O(t^{2-c})\leq1/t$ for all
sufficiently large $t$, proving the first item.

For the second item, suppose towards contradiction that for some
$c>0$ and infinitely many $t$, the $(t^c,t^{-c})$ insertional
pseudoentropy gap is less than $\tfrac\gamma2 t^{2\alpha}$.
Set $\delta=\tfrac\gamma2 t^{2\alpha}+t^{-c}$, so the insertional
pseudoentropy is at most $H(L)+\delta-t^{-c}$.
Part 2 of \Cref{thm:insertional-characterization} gives a stateless
insertional sampler of size
$\poly(t^c,2t,2^\ell,t^c)=2^{O(t)}$ and KL error at most
$\delta$. Since $2\alpha>1$, for all sufficiently large $t$ we have
$\delta<\gamma t^{2\alpha}-2t$, contradicting
\Cref{thm:sm-insertional-separation}.
\end{proof}

\section{Consequences}
\label{sec:consequences}
This section explores consequences of our results. In \Cref{sec:consequences-crypto},
we derive limitations on sequential generation of cryptographic distributions with limited private state.
In \Cref{subsec:insertional-vs-ltr-generation}, we explain how our results
motivate insertion-based models by showing the added expressive power of
choosing where to generate each token.

\subsection{Cryptographic hardness for sequential sampling}
\label{sec:consequences-crypto}
Our results show that small-state sequential samplers cannot do many standard cryptographic tasks. We give two examples: (1) a small-state sequential sampler cannot sample a pseudorandom generator distribution, and (2) a small-state sequential sampler cannot sampler fresh public-key ciphertexts even given the public key. This shows that, in order to compute cryptographic primitives, autoregressive models require either external mechanisms or a large state. On the flip-side, we show that sequential samplers cannot invoke public-key steganography. Thus, agents with no pre-shared secrets and a small state cannot covertly communicate with each other. This may be useful for security and alignment, in cases where an autoregressive agent  is not trusted.

In order to describe consequences of our results to sequential generation of cryptographic primitives, we first derive a general consequence of
\Cref{thm:sequential-characterization,thm:state-elimination-KL} for
negligible sampling error, allowing logarithmic private state. We use $n$ as the asymptotic parameter throughout this subsection.

\begin{corollary}[General consequence for samplers with logarithmic state]
\label{cor:stateless-general-consequence}
Let $(W,Y)$ be jointly distributed and indexed by $n$. Assume:
\begin{enumerate}[noitemsep]
    \item $W$ has polynomial length, and $Y$ consists of $\poly(n)$ items over a $\poly(n)$-size alphabet.
    \item $Y$ has a noticeable nonuniform next-block pseudoentropy
    gap given $W$.
\end{enumerate}
Then no sequential sampler $C$ with $\poly(n)$-size circuits and
$O(\log n)$ private bits satisfies
\[
\kldiv{(W,Y)}{(W,C(W))}=\negl(n),
\]
even given $W$ and the entire visible prefix at every step.
\end{corollary}
\begin{proof}
Suppose towards contradiction that such a sampler $C$ exists, and
write $k=O(\log n)$ for its state length. Let $N\geq1$ be the number
of tokens of $Y$, each of length $\ell$. Put
$P=(W,Y)$, $Q=(W,C(W))$, and $e=\kldiv{P}{Q}=\negl(n)$.
Choose $d>0$ so that the pseudoentropy gap is at least $n^{-d}$
for all sufficiently large $n$, and set $\rho=n^{-d}/8$.
We remove the private state, control the resulting error, and then
apply the stateless characterization.
\begin{enumerate}
    \item \emph{Remove the private state.}
    Apply \Cref{thm:state-elimination-KL} with accuracy $\rho$.
    Its construction allows visible input $W$: use the same $w$ in
    every simulated invocation of $C$, and average the KL bound over
    $W$. The resulting stateless sampler $D$ has polynomial size,
    since $2^k,2^\ell,1/\rho,N,|W|$ are polynomially bounded.
    Writing $R=(W,D(W))$, we have $\kldiv{Q}{R}\leq\rho$.

    \item \emph{Compare with the target distribution.}
    Both $C$ and $D$ use at most $r=\poly(n)$ fair random bits
    over the entire run. Thus every positive conditional output
    probability, given $W=w$, is at least $2^{-r}$.
    Finite KL-divergence gives
    $\supp(P)\subseteq\supp(Q)\subseteq\supp(R)$.
    On $\supp(Q)$, put
    $F(w,y)=\log(\Pr[C(w)=y]/\Pr[D(w)=y])$, and set $F=0$ elsewhere.
    Then $|F|\leq r$.
    By \Cref{fact:pinsker}, $\nu:=\statDist(P,Q)=\negl(n)$.
    Since all three distributions have the same $W$ marginal,
    \begin{align*}
    \kldiv{P}{R}
    &=e+\kldiv{Q}{R}+\E_P F-\E_Q F\\
    &\leq e+\rho+2r\nu
    =\rho+\negl(n)\leq2\rho
    \end{align*}
    for all sufficiently large $n$.

    \item \emph{Apply the pseudoentropy lower bound.}
    Part 1 of \Cref{thm:sequential-characterization} allows visible
    input $W$: replace $(I,Y_{<I})$ by $(W,I,Y_{<I})$, condition the
    entropies on $W$, and include $|W|$ in the polynomial size bound.
    Set $\delta=4\rho$, $\sigma=\rho$, $\eta=\rho/N$,
    $t=n^c$, and $\epsilon=n^{-c}$.
    The denominator in the size bound is polynomial in $n$
    independently of $c$, while the numerator is
    $\min\{n^{cc_0},\rho n^c/N\}$.
    Choose $c$ large enough to cover the size of $D$ and ensure
    $n^{-c}\leq\rho$. The pseudoentropy hypothesis then gives
    $(n^c,n^{-c})$ next-block pseudoentropy given $W$
    at least $H(Y\mid W)+8\rho-n^{-c}\geq H(Y\mid W)+4\rho$.
    The theorem yields $\kldiv{P}{R}>4\rho-\rho-N(\rho/N)=2\rho$,
    contradicting the preceding bound.
\end{enumerate}
\end{proof}

We now describe several corollaries for various cryptographic consequences of the above corollary. 

\begin{corollary}[Pseudorandom generation]
\label{cor:no-stateless-prg}
Let $G:\zo^{k(n)}\to\zo^n$ be a pseudorandom generator secure against
nonuniform polynomial-size circuits. No sequential bit sampler $C$ with $\poly(n)$-size circuits and $O(\log n)$ private bits satisfies
\[
\kldiv{G(U_{k(n)})}{C(1^n)}=\negl(n).
\]
The same holds for any apriori fixed ordering, rather than just left-to-right.
\end{corollary}

A \emph{public-key encryption scheme} generates a public key $\mathsf{pk}$
and a secret key $\mathsf{sk}$. Anyone can encrypt a message using
$\mathsf{pk}$; decryption using $\mathsf{sk}$ recovers it.
\emph{IND-CPA security} requires indistinguishable encryptions of any two
equal-length messages chosen efficiently after seeing $\mathsf{pk}$.
\emph{Public-key steganography} additionally requires encodings to be
indistinguishable from a prescribed cover distribution, such as ordinary
text. The following results concern this public-key setting, without a
pre-shared secret between sender and receiver.

\begin{corollary}[Public-key encryption]
\label{thm:no-stateless-pke}
Let a polynomial-time public-key encryption scheme have security
parameter $n$. Assume:
\begin{enumerate}[noitemsep]
    \item IND-CPA security against nonuniform polynomial-size circuits
    and negligible correctness error.
    \item Ciphertexts consist of a fixed polynomially bounded number
    of $O(\log n)$-bit tokens.
\end{enumerate}
For any two distinct equal-length messages of polynomial length, their
ciphertext distributions cannot both be sampled with negligible
KL-divergence by sequential samplers with $\poly(n)$-size circuits and
$O(\log n)$ private bits.
The public key, message, and visible prefix are available at every step;
divergence is measured jointly with the public key.
\end{corollary}

\begin{corollary}[Public-key steganography]
\label{cor:no-stateless-steganography}
Let a public-key steganography scheme run in polynomial time, with
security parameter $n$. Fix two distinct messages of polynomial length.
Assume:
\begin{enumerate}[noitemsep]
    \item Negligible decoding error; both encodings consist of the same
    fixed polynomially bounded number of $O(\log n)$-bit tokens.
    \item Both encodings are computationally indistinguishable from the
    same cover distribution, jointly with the public key, against
    nonuniform polynomial-size circuits.
\end{enumerate}
The two encoding distributions cannot both be sampled with negligible
KL-divergence by sequential samplers with $\poly(n)$-size circuits and
$O(\log n)$ private bits.
The public key, message, and visible prefix are available at every step;
divergence is measured jointly with the public key.
\end{corollary}

\begin{proof}[Proof of the three corollaries]
For the PRG, we describe it for left-to-right ordering, but the same argument holds for any fixed known ordering up to applying permutation. Put $Y=G(U_{k(n)})$. For uniform $I\in[n]$ and an
independent uniform bit $B$, the distributions $(I,Y_{<I},Y_I)$ and
$(I,Y_{<I},B)$ are computationally indistinguishable: replacing $Y$
by $U_n$ on both sides makes them identical. Thus $Y$ has next-bit
pseudoentropy $n$, whereas $H(Y)\leq k(n)<n$.
\Cref{cor:stateless-general-consequence} applies with constant visible input.

For the remaining claims, let $P_b=(W,Y^b)$ be the two encoding
distributions, including their common public key or setup $W$.
Write $N$ for the number of tokens and $u=2^\ell$ for the alphabet size,
where $\ell$ is the token length.
They are computationally indistinguishable and satisfy
$\statDist(P_0,P_1)\geq1-\negl(n)$.
For encryption and steganography, correctness gives the distance bound:
appending the secret key preserves distance, since it is independent of
the encoding given the public key. Steganographic security gives
indistinguishability through the common cover. 

Suppose towards contradiction that both samplers exist. We distinguish
$P_0$ from $P_1$ in three steps.
\begin{enumerate}
    \item \emph{Remove the private state.}
    Put $\alpha=1/(100N)$. Apply \Cref{thm:state-elimination-KL}
    with KL error $\alpha^2/4$, keeping $W$ visible.
    By \Cref{fact:pinsker} and the assumed negligible sampling error,
    the resulting polynomial-size stateless samplers $D^b$ satisfy
    $\statDist(P_b,R_b)\leq\alpha$, where $R_b=(W,D^b(W))$.
    Hence $\statDist(R_0,R_1)\geq1/3$ for all sufficiently large $n$.

    \item \emph{Find a differing transition.}
    Replacing the circuits of $D^0$ by those of $D^1$ one at a time
    gives an index $i$ such that, for $X=(W,D^0(W)_{<i})$ and
    $A_b\gets D_i^b(X)$ with fresh coins,
    $d:=\statDist((X,A_0),(X,A_1))\geq1/(3N)$.
    Each hybrid uses the message-$0$ prefix and a common message-$1$
    suffix, which cannot increase distance.
    Yet every polynomial-size test distinguishes these pairs with
    advantage at most $4\alpha+\negl(n)$: replacing the
    $R_0$ prefix by the $R_1$ prefix through $P_0,P_1$ before invoking
    $D_i^1$ costs
    $2\alpha+\negl(n)$, and comparing the two honest prefixes
    including token $i$ costs the same amount.

    \item \emph{Test the transition probabilities.}
    On input $(x,a)$, draw $M=\lceil576uN^2\rceil$ samples from each
    $D_i^b(x)$, and accept if $a$ occurs more often in the first batch.
    Write $p_b$ for the true token distribution and $\widehat p_b$
    for its empirical distribution. By Cauchy--Schwarz,
    $\E\|\widehat p_b-p_b\|_1\leq\sqrt{u/M}\leq1/(24N)$.
    Choosing the larger empirical probability loses at most
    $\|\widehat p_0-p_0\|_1+\|\widehat p_1-p_1\|_1$ relative to
    the optimal token test. Averaging over $X$, our test therefore has
    advantage at least $d-1/(12N)\geq1/(4N)$, contradicting the
    preceding upper bound.
\end{enumerate}
The index $i$ is nonuniform advice, and all circuits have polynomial size.
The test's random coins can be fixed by averaging.

The transition-testing argument also gives an alternative proof of the
PRG claim. Take $W$ constant, $N=n$, $\ell=1$, and $u=2$, with
$P_0=G(U_{k(n)})$ and $P_1=U_n$. These distributions are computationally
indistinguishable, while the support-size bound gives
\[
\statDist(P_0,P_1)\geq1-2^{k(n)-n}\geq1/2.
\]
The sampler for $P_1$ draws independent uniform bits exactly, so a
hypothesized sampler for $P_0$ supplies both samplers needed above.
With $\alpha=1/(100n)$, the first step gives
$\statDist(R_0,R_1)\geq1/2-2\alpha\geq1/3$ for all sufficiently large $n$.
The remaining two steps, applied to this pair of distributions, yield
the same contradiction. As above, any fixed known ordering is handled
by permuting the coordinates.
\end{proof}

\subsection{Insertional versus Left-to-Right Token Generation}
\label{subsec:insertional-vs-ltr-generation}

\citet{DBLP:conf/icml/SternCKU19} introduce an insertion
framework and implement it with the \emph{Insertion Transformer}. Given
an input sequence $x$ and the current partial output $w$, called the
\emph{canvas}, the model assigns a joint distribution $p(a,j\mid x,w)$
to the next token $a$ and its insertion position $j$. The decoder's hidden
states are recomputed after each insertion, so earlier computation leaves
no private state beyond the canvas~\cite[Section~4.5]{DBLP:conf/icml/SternCKU19}.

To connect this framework to ours, fix the input $x$ and the output
length, and consider sampling one insertion per step with fresh coins.
This is a sampling version of their framework; their decoding procedures
choose maximum-probability actions, either serially or in
parallel~\cite[Section~5]{DBLP:conf/icml/SternCKU19}.
Allowing arbitrary randomized circuits for these transitions gives exactly
our stateless insertional samplers, up to polynomial overhead
(\Cref{def:statelessInsertionalSampler}). Indeed, a token and position
determine the next canvas; conversely, a one-token extension determines
a valid insertion by locating a token whose deletion recovers the old
canvas. Restricting every insertion to the rightmost position gives
stateless sequential sampling.

Our separation now gives a computational motivation for insertion-based
models. Assuming one-way functions secure against polynomial-size circuits
exist, \Cref{thm:insertional-vs-sequential} shows that efficient insertional
generation can exactly sample distributions over binary strings that
efficient left-to-right generation cannot sample even with small KL error.
The advantage already appears with one insertion per step: the generator
can produce information needed for later decisions before filling in
earlier output positions. Thus choosing where to write can expand what
is efficiently generatable without adding private memory or requiring
parallel decoding. Indeed, as we
have shown in \Cref{cor:no-stateless-prg,prop:insertionally-samplable-prg}, while PRGs are hard to sample for sequential samplers (for any fixed ordering), even with
O(log n) bits of state, this is not the case for insertional samplers.

Moreover, the notion of insertional generation does not seem to be captured by the monotone compression notion of \cite{MMP26}. This gives further motivation to explore this model of computation (as well as models that allow deletions or substitutions).


\subsection*{Acknowledgements}
We thank Adam Kalai for discussions that initiated this research. 
G.A.,~A.R.,~and P.L.~are supported by the European Research Council (ERC) under the EU’s Horizon 2020 research and innovation programme (Grant agreement No. 101019547) and by a Cariplo CRYPTONOMEX grant. The majority of this work was done while G.A. was a postdoctoral fellow at Bocconi University.
G.M.~is supported by the European Research Council through an ERC Starting Grant (Grant agreement No.~101077455, ObfusQation) and by the Deutsche Forschungsgemeinschaft (DFG, German Research Foundation) under Germany's Excellence Strategy - EXC 2092 CASA – 390781972.  A.B. and L.T. are supported by an NSERC Discovery Grant.

\subsection*{AI use statement}
The core research ideas are due to the authors. We used large
language models, specifically ChatGPT 6 Astra, to assist with writing and revising the manuscript,
including proofs, and to help check mathematical arguments for gaps
and errors. The authors take full responsibility for the final
content of the paper, including its claims, proofs, and references.

\bibliographystyle{alpha_nat}
\bibliography{abbrev3,crypto,custom}
\fi

\appendix

\end{document}